\documentclass{article}

\PassOptionsToPackage{numbers, compress}{natbib}

\usepackage[main, final]{neurips_2026}

\usepackage[utf8]{inputenc} 
\usepackage[T1]{fontenc}    
\usepackage{hyperref}       
\usepackage{url}            
\usepackage{booktabs}       
\usepackage{amsfonts}       
\usepackage{amsmath}
\usepackage{amssymb}
\usepackage{nicefrac}       
\usepackage{microtype}      
\usepackage{xcolor}         
\usepackage{graphicx}
\usepackage{eso-pic}

\title{Neural Modal Decomposition: Architectural Priors from Observables}

\author{%
  Juho Park, Kaushik Sengupta \\
  Princeton University \\
  \texttt{\{jp3327, kaushiks\}@princeton.edu} \\
}

\begin{document}
\AddToShipoutPictureFG*{%
  \AtPageUpperLeft{%
    \put(\LenToUnit{1.5in},\LenToUnit{-72pt}){%
      \includegraphics[height=28pt]{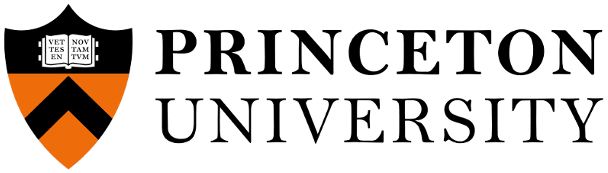}}}}

\maketitle

\begin{abstract}
Many engineering building blocks behave as multi-port linear time-invariant systems. RF cavities, photonic devices, and superconducting quantum chips, despite their different underlying physics, all share a common mathematical structure for their port-level response. Each entry of the response matrix is a sum of contributions from a small number of intrinsic resonant modes, the pole-residue form. A model capable of predicting such responses for arbitrary geometries and arbitrary port
configurations, while simultaneously extracting the underlying eigenmode structure, could therefore offer a design principle potentially applicable across all these domains. We propose a neural framework that learns this modal decomposition end-to-end, supervised only by system-level observables and without supervising the modal parameters themselves. The architecture decomposes into a port-independent pole predictor and two port-dependent coupling predictors whose outputs are combined entry-wise, separating intrinsic from port-dependent features. This factorization yields a single trained model that generalizes to port counts unseen during training, whereas direct regression needs a separate model for each port count, with an output size that grows as $\mathcal{O}(N^2)$. Although the modal parameters are never supervised, the freely parameterized poles converge to the physical resonances, in agreement with poles extracted by the AAA rational approximation algorithm. We instantiate the framework in radio-frequency electromagnetic surrogate modeling. A model trained only on 2-port data accurately predicts $N$-port responses unseen during training.
\end{abstract}

\section{Introduction}
\label{sec:intro}

\begin{figure}[!htbp]
  \centering
  \includegraphics[width=0.97\textwidth]{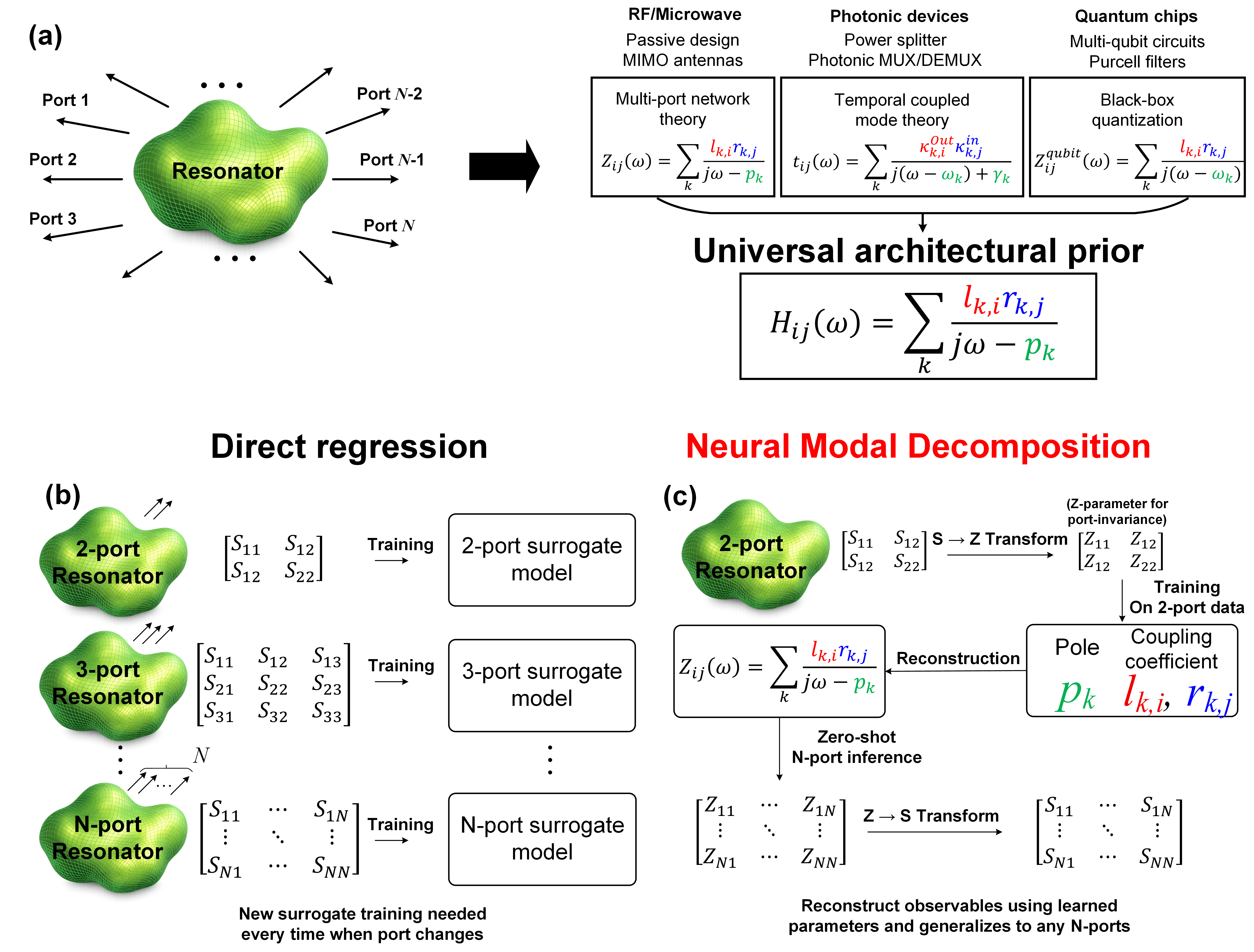}
  \caption{\textbf{A shared architectural prior across physical domains, and comparison between direct regression and NMD.} (a) Multi-port LTI systems in RF/microwave engineering, photonic devices, and superconducting quantum chips share the pole-residue form. (b) Direct regression on $S$-parameters needs a separate dataset and model for each port count $N$. (c) NMD converts the 2-port $S$ training data to the impedance $Z$, which keeps the same poles when ports are added or removed (Section~\ref{sec:method}), learns $p_k$ and the per-port couplings $l_{k,i}, r_{k,j}$, and assembles $Z$ for any $N$ ports zero-shot before converting back to $S$.}
  \label{fig:overview}
\end{figure}

A wide range of engineering building blocks share the same physical character. They are linear time-invariant (LTI) multi-port systems whose response is dominated by a few intrinsic resonant modes. RF cavities and passive components \citep{karahan2024deepmultiport,karahan2024deepantenna}, photonic devices such as power splitters and wavelength demultiplexers \citep{tahersima2019deep,piggott2015inverse}, and the linear electromagnetic environment of superconducting quantum chips \citep{nigg2012black,arute2019quantum,jeffrey2014fast} all admit a common port-level description (Figure~\ref{fig:overview}a). Each entry of the multi-port response matrix can be written as
\begin{equation}
H_{ij}(\omega) = \sum_{k} \frac{l_{k,i}\, r_{k,j}}{j\omega - p_k},
\end{equation}
a sum over intrinsic modes with port-independent poles $p_k$ and port-dependent coupling coefficients $l_{k,i}, r_{k,j}$. This is the \emph{pole-residue form}, the standard form of the response of an LTI multi-port system with a discrete set of modes. Within each domain it appears under different names (multi-port network theory, temporal coupled-mode theory, black-box quantization), but the algebra is the same.

\paragraph{Why observables, not fields.} For these systems, the practical quantity of interest at design time is the system-level observable. Examples include the impedance matrix in microwave engineering, the transmission matrix in photonics, and the multi-port impedance at qubit ports in circuit quantum electrodynamics (cQED). The internal field is an intermediate representation, expensive to simulate or measure and rarely the target of design itself. Field-based neural surrogates such as PINNs \citep{raissi2019physics} and neural operators \citep{li2021fno,lu2021deeponet} provide powerful tools when the spatial field is itself the quantity of interest. For multi-port system design, however, where the response matrix is the direct design target, observable-based modeling offers a complementary and more compact route.

\paragraph{Modal decomposition gives port-count generalization.} A useful consequence of the pole-residue form is that its dimensionality does not depend on the number of ports observed. The poles are intrinsic to the system and shared across all port configurations, and only the per-port couplings need to be evaluated for each port. A surrogate that outputs modal parameters can therefore predict the response at any port count, including port numbers larger than those seen in training. Direct regression on the response matrix, by contrast, scales output dimensionality as $\mathcal{O}(N^2)$, requires a separately collected dataset for each port count, and cannot generalize to new port configurations (Figure~\ref{fig:overview}b,c).

\paragraph{Why this has been hard.} Despite this structural benefit, learning modal decompositions with neural networks has remained difficult. Direct supervision of pole-residue parameters \citep{feng2015parametric,feng2017parametric,zhang2021advanced} requires first running rational fitting algorithms such as Vector Fitting \citep{gustavsen2002rational} on the training data and then training a network to regress these parameters. This pipeline inherits well-known pathologies. Poles can change in number across samples, residues admit sign and phase ambiguities, and spurious modes appear when the rational fit is over-parameterized. The deeper issue is that pole-residue parameters are poor regression targets. The same response can be written with many parameter sets (the poles are unordered quantities, for example), and the parameters can change abruptly when the geometry changes slightly.

\paragraph{Our approach.} We perform modal decomposition through architectural design, supervised only by system-level observables and never supervising the modal parameters themselves. The forward computation is a simple pole-residue sum, and the network's three modules predict the poles and coupling coefficients. A cavity encoder produces a representation of the structure alone, a pole predictor maps it to $K$ complex poles, and a port-dependent coupling predictor is queried separately for the left and right ports of each entry to produce $l_{k,i}$ and $r_{k,j}$. Loss is computed on observables. The factorization $l_{k,i} r_{k,j}$ makes the loss invariant to a discrete sign symmetry of the residues, and a gauge term fixes the remaining phase ambiguity across port queries. Although the modal parameters are never supervised, the freely parameterized poles converge to the physical resonances, which we compare with poles extracted by the AAA rational approximation algorithm \citep{nakatsukasa2018aaa}.

\paragraph{Contributions.} We summarize our contributions as follows.
\begin{itemize}
    \item We propose a neural framework that learns the multi-port pole-residue decomposition end-to-end from system-level observables, without supervising modal parameters. The architecture factors into a port-independent pole predictor and a port-dependent coupling predictor combined entry-wise, with an anchor-gauge term that resolves the sign degeneracy across port queries. Because photonic and superconducting quantum devices share the same pole-residue form, the prior can potentially be extended to these domains.

    \item We instantiate the framework in radio-frequency electromagnetic surrogate modeling. A model trained only on 2-port data accurately predicts 3-, 4-, and 5-port responses unseen during training, achieving zero-shot $N$-port generalization.

    \item We show that, although the modal parameters are never supervised, the freely parameterized poles converge to the physical resonances, in agreement with poles extracted by the AAA rational approximation algorithm \citep{nakatsukasa2018aaa} and with the exact poles of RLC networks.
\end{itemize}

\section{Related Work}
\label{sec:related}

\paragraph{Direct supervision of modal parameters.} The Neuro-Transfer-Function line of work \citep{feng2015parametric, feng2017parametric, zhang2021advanced} extracts pole-residue parameters from training data using Vector Fitting \citep{gustavsen2002rational}, then supervises a network to regress these parameters. Pole-residue variants suffer from pole tracking and order-changing issues, and rational variants face severe numerical sensitivity at high orders \citep{zhang2021advanced}. Subsequent work has been substantially devoted to patching these issues \citep{zhang2020novel, zhang2021advanced}, but the underlying difficulty remains, since the same response can be written with many pole-residue parameter sets and the parameters can change abruptly across samples. Our framework treats poles and residues as free parameters of the forward computation, supervised only implicitly through observables.

\paragraph{Physics-informed neural networks (PINNs).} PINNs \citep{raissi2019physics} regularize predictions with PDE-residual losses, and neural operators \citep{li2021fno, lu2021deeponet} learn function-to-function mappings between input and output fields. These approaches are powerful when the spatial field is the quantity of interest, but they target the high-dimensional field itself, with modal characteristics being extracted afterwards. PINNs themselves are mesh-free, but the training field data needs mesh-based solvers. Our contrast is therefore between field-based and observable-based modeling. Our framework targets a complementary problem class (system-level observables) and exploits the algebraic structure of the observable to obtain compact port-level outputs and interpretable modal parameters.

\paragraph{Spectral learning in dynamical systems.} DeepKoopman \citep{lusch2018deep} and follow-ups \citep{takeishi2017learning, otto2019linearly, yeung2019learning} train architectures with linear-dynamics bottlenecks to recover spectral parameters such as Koopman eigenvalues without direct supervision. They share with us the principle that spectral parameters can emerge from architectural design under observable-level losses, but operate on time-domain trajectories of single systems rather than on a parametric family of multi-port responses, target scalar observables rather than the matrix-valued multi-port observable whose submatrix structure we exploit, and rely on prediction accuracy alone as evidence of spectral correctness, without independent cross-validation against true physical modes, which is a check we provide via AAA algorithm.

\paragraph{Structure-constrained architectures.} Our work belongs to a broader line showing that hard-coding known structure into an architecture improves generalization. Projected Neural ODEs \citep{pal2025semiexplicit}, for example, are given a single dynamical system together with a conservation law known in advance as an algebraic constraint, and project every predicted state onto it so that long simulations do not drift. The two settings differ in what is known. In ours nothing about the modes is known in advance. The network is told only that the response has the form of a pole-residue sum, and the number of modes, the pole locations, and the port couplings are discovered during training across a family of geometries. Projected Neural ODEs enforce a known invariant to extrapolate in time, whereas NMD uses an assumed form to discover unknown structure and extrapolate in port count.

\section{Method}
\label{sec:method}

\subsection{Pole-residue form as architectural prior}

For an LTI multi-port physical system with $N$ input/output channels, the matrix-valued transfer function $H(\omega) \in \mathbb{C}^{N \times N}$ admits the pole-residue decomposition
\begin{equation}
\label{eq:pole-residue}
H_{ij}(\omega) = \sum_{k=1}^{K} \frac{l_{k,i}\, r_{k,j}}{j\omega - p_k},
\end{equation}
where $\{p_k\}_{k=1}^K \subset \mathbb{C}$ are complex poles intrinsic to the system and $\{l_{k,\cdot}, r_{k,\cdot}\}_{k=1}^K \subset \mathbb{C}^N$ are coupling vectors describing how each mode couples to each port. Reciprocal systems satisfy $l_k = r_k$ as a special case. We keep the general form so the framework applies equally to non-reciprocal systems such as non-reciprocal cQED architectures. The framework applies to LTI systems whose response over the band of interest is dominated by a discrete set of resonant modes, so that a finite number $K$ of poles suffices.

In our radio-frequency electromagnetic experiments, we adopt the impedance matrix $Z(\omega) \equiv H(\omega)$ as the working representation. It is related to the scattering matrix $S(\omega)$ by the standard impedance-to-scattering transformation $S = (Z - Z_0 I)(Z + Z_0 I)^{-1}$ with reference impedance $Z_0 = 50\,\Omega$. We use $H(\omega)$ in this section to emphasize that the prior itself is system-agnostic. Experiments adopt the RF-specific $Z, S$ notation.

\paragraph{Submatrix consistency.} A central algebraic property we exploit is that the impedance representation is \emph{consistent under port-subset restriction}. For any port subset $\mathcal{P} \subset \{1,\dots,N\}$, the impedance matrix obtained by physically open-terminating the unselected ports equals the submatrix $Z_{\mathcal{P},\mathcal{P}}(\omega)$ of the original. The same poles $\{p_k\}$ describe the system regardless of how many ports are observed, and only the coupling vectors are restricted to the selected ports. This motivates an architecture in which pole prediction is decoupled from port observation. Importantly, submatrix consistency is \emph{not} universal. The scattering matrix $S$ does not satisfy it because $S$ is defined with every port terminated in a reference load, and those loads absorb energy and alter the field. Adding or removing a port therefore changes every entry of $S$, not only the entries involving that port. Choosing the impedance over the scattering representation is what makes pole prediction independent of the port set.

The deeper reason is that impedance corresponds to non-invasive probing of an internal system. Adding or removing measurement ports does not alter internal dynamics, only which channels are observed. Any LTI multi-port system admits the resolvent-style representation $H_{ij}(\omega) = \phi_i^{\mathit{T}} (j\omega I - L)^{-1} \psi_j$, with $L$ an internal operator independent of the port set and $\phi_i, \psi_j$ port-specific input/output mappings \citep{kailath1980linear}. Restricting attention to a port subset selects a subset of these mappings while leaving $L$ unchanged, yielding exactly the submatrix of $H$. The same characterization holds across LTI domains. Examples include the impedance and admittance matrices in microwave theory \citep{pozar2011microwave}, the receptance matrix in structural dynamics \citep{ewins2009modal}, multi-port acoustic impedance matrices in branched duct networks \citep{munjal1987acoustics}, and multi-channel resolvents in quantum scattering \citep{taylor2012scattering}. In this work we instantiate the framework in radio-frequency electromagnetics, leaving extensions to other domains for future work.

\subsection{Architecture}
\label{sec:architecture}

We instantiate Equation~\ref{eq:pole-residue} as a neural network composed of three modules (Figure~\ref{fig:architecture}c). The network takes as input the cavity geometry and a pair of port locations $(x_i, x_j)$, and outputs the entries of the multi-port response. Layer-level specifications, dimensions, and training details are deferred to Appendix~\ref{app:training}.

\paragraph{CavityNet (structure encoder).} CavityNet $\Phi_{\text{cav}}(g)$ takes the cavity geometry as a binary pixel pattern and produces a sequence of intrinsic cavity tokens, using a convolutional backbone followed by tokenization. CavityNet does not see port locations and captures only properties intrinsic to the cavity itself.

\paragraph{PoleNet.} PoleNet $\Phi_{\text{pole}}$ maps the cavity tokens to a set of $K$ complex poles. We parameterize each pole as
\begin{equation}
\label{eq:pole-param}
p_k = -\gamma_k + j\,\omega_k, \qquad \gamma_k > 0,
\end{equation}
where $\omega_k$ is the resonance frequency of mode $k$ and $\gamma_k$ is its damping rate. The damping rate is constrained positive (via a softplus on a raw network output) so that the pole always lies in the left half-plane and the system is stable by construction. Throughout the paper we plot poles in the complex frequency plane $\omega_p = -j p_k = \omega_k + j\gamma_k$, so that the horizontal axis is the resonance frequency and the vertical axis is the damping rate (Appendix~\ref{app:polenet} gives the explicit conversion). PoleNet depends only on cavity tokens, reflecting the physical fact that poles are intrinsic to the cavity and independent of port placement.

\paragraph{AmpNet (left and right).} AmpNet maps the cavity tokens together with the location of a queried port into per-pole coupling entries, with two independent heads producing the left coupling $l_{k,i}$ and the right coupling $r_{k,j}$. The port location $(r_i, c_i)$ is encoded with a hybrid scheme. The row index takes only two values in our setup, the top and bottom edges of the cavity. A sinusoidal encoding of a two-valued input returns nearly identical feature vectors for both values, leaving the network almost no signal to distinguish them, so we use a learned embedding with one vector per row instead. The column index spans 18 distinct values and admits a standard sinusoidal positional encoding. The row and column encodings are concatenated and projected into a single vector representing the queried port. This vector attends to the cavity tokens produced by CavityNet, so that the port representation is conditioned on the geometry, and two linear heads then output the magnitude and phase of $l_{k,i}$ and $r_{k,j}$ for each pole.

\paragraph{Synthesis.} For a port pair $(i, j)$ the impedance entry $Z_{ij}(\omega)$ is computed entry-wise by Equation~\ref{eq:pole-residue} with $H \equiv Z$. The full $N{\times}N$ impedance matrix is assembled by querying all $N$ ports through both AmpNet heads, and is then converted to scattering parameters via the bilinear transformation. Because each port is queried independently, the synthesis operates at any port count without architectural change.

\subsection{Sign invariance and the gauge-fixing loss}

The pole-residue form admits a discrete sign symmetry. Replacing $(l_k, r_k) \to (-l_k, -r_k)$ leaves the entry-wise product $l_{k,i} r_{k,j}$ invariant, so $Z$ and the reconstruction loss are unchanged. This gauge ambiguity is one of the reasons why direct supervision of residues is ill-posed. Any modal supervision target can be matched by $2^K$ different network outputs, all representing the same residue. Our $l_{k,i} r_{k,j}$ factorization makes the reconstruction loss automatically invariant to this symmetry, eliminating a primary source of training instability.

The reconstruction loss computed on $Z$ and $S$ is therefore identical for all $2^K$ sign-equivalent solutions, which implies a practical hazard. Residues may drift along these sign-equivalent solutions during training, and nothing prevents the gauge from being chosen inconsistently across different port queries within the same sample. Such inconsistency can corrupt the assembled outer product across the $N \times N$ matrix. Empirically, removing the gauge-fixing term leaves zero-shot 3-, 4-, and 5-port accuracy essentially unchanged, while keeping it reduces the 2-port $S$-parameter MAE by roughly 16\%, so we regard it as a lightweight regularizer for in-band fidelity (Appendix~\ref{app:ablation} ablates the gauge term and $K$).

To fix a canonical gauge, we designate the bottom edge (row index $r_{\mathrm{anchor}} = 17$) as the anchor edge. For every queried port that lies on this edge we impose, on each predicted modal amplitude $a_k$ (drawn from either the left or right AmpNet head), the per-mode penalty
\begin{equation}
\label{eq:gauge-fix}
\ell_{\text{gauge}}(a_k) = 1 - \frac{\mathrm{Re}(a_k)}{|a_k| + \varepsilon_{\text{anchor}}},
\end{equation}
which is minimized when $a_k$ has phase zero on the positive real axis ($\varepsilon_{\text{anchor}} = 10^{-6}$ stabilizes the phase computation and is distinct from the log-magnitude stabilizer $\varepsilon$ in Equation~\ref{eq:loss}). The total gauge loss aggregates this penalty over both AmpNet heads and over both queried ports of a sample, and is applied only when the queried port row equals $r_{\mathrm{anchor}}$. We anchor only at $r_{\mathrm{anchor}}$ rather than at every port, since the sign ambiguity (one choice of sign per mode, $2^K$ in total) is removed once a single reference is fixed. Anchoring additional ports such as those on the top edge would over-constrain the network, because the relative amplitudes and phases between top and bottom edge couplings carry the actual mode-shape information that the network must be free to learn. The full aggregated form, including masks, is given in Appendix~\ref{app:gauge}.

\subsection{Training}
\label{sec:training}
We instantiate the framework in the RF setting, where the relevant observables are the impedance matrix $Z(\omega)$ and the scattering matrix $S(\omega)$. The total training loss combines two observable-level reconstruction terms with the gauge-fix term,
\begin{equation}
\label{eq:loss}
\mathcal{L} = \underbrace{\| S_{\text{pred}} - S_{\text{gt}} \|}_{\mathcal{L}_S} \;+\; \lambda_Z \underbrace{\big\| \log(|Z_{\text{pred}}|+\varepsilon) - \log(|Z_{\text{gt}}|+\varepsilon) \big\|}_{\mathcal{L}_{Z,\log}} \;+\; \lambda_g\, \mathcal{L}_{\text{gauge}}.
\end{equation}
Here $\|\cdot\|$ denotes the mean absolute error, averaged over all matrix entries, all frequencies, and, for complex quantities, the real and imaginary parts. We supervise on both $S$ and $Z$ representations because they are complementary in their conditioning. $S$ is well-conditioned away from resonance, while $Z$ has a wide dynamic range near resonance that the log-magnitude form $\mathcal{L}_{Z,\log}$ handles stably ($\varepsilon$ is a small numerical stabilizer). \emph{No supervision on $\{p_k, l_{k,i}, r_{k,j}\}$ is provided.} Training data consists exclusively of $N=2$ channel observations. Generalization to $N \ge 3$ is evaluated zero-shot at inference time.

\section{Experiments}
\label{sec:experiments}

\begin{figure}[!htbp]
  \centering
  \includegraphics[width=0.88\textwidth]{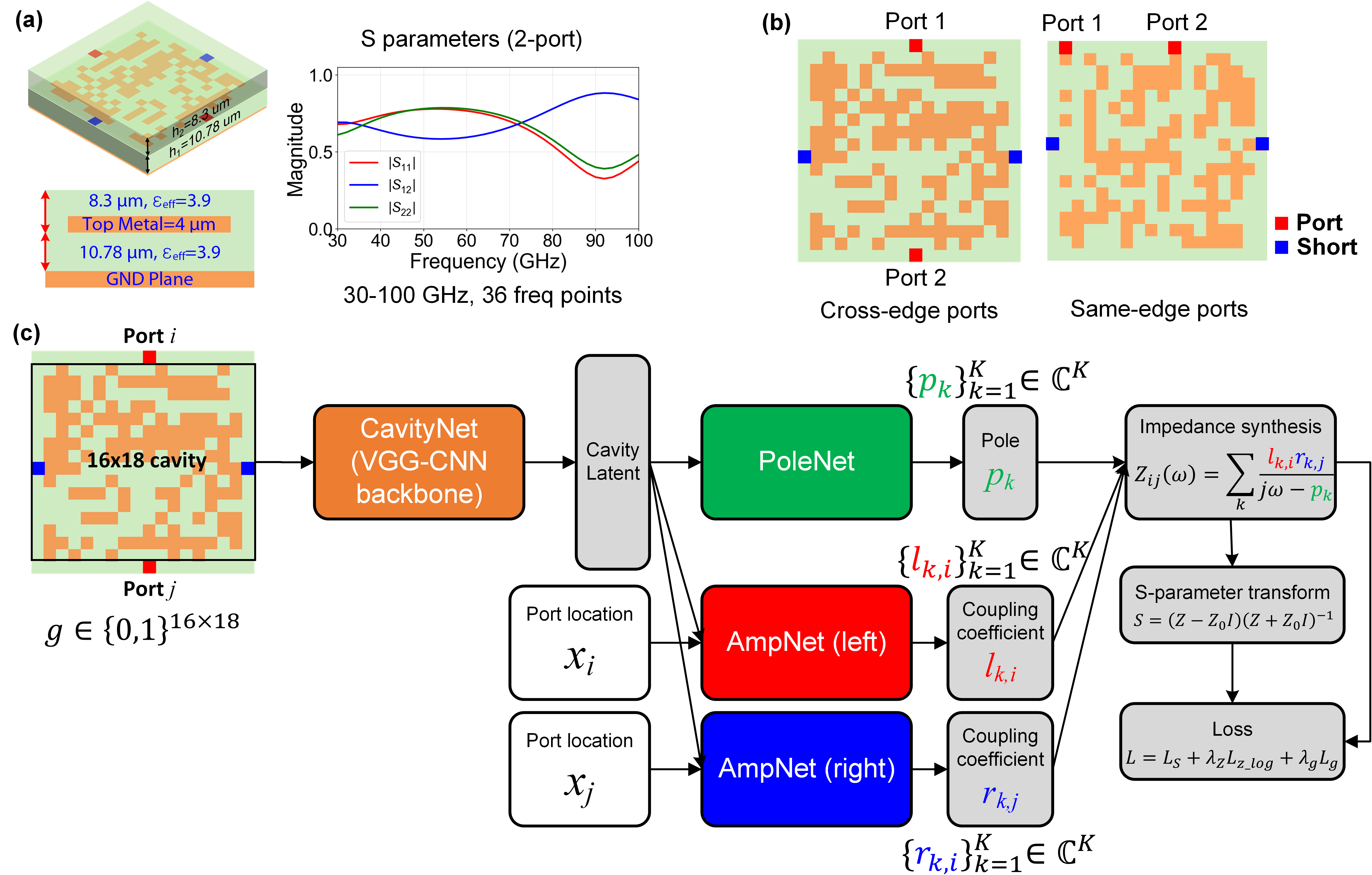}
  \caption{\textbf{Dataset, port arrangements, and architecture.} (a) Cavity stack-up. A patterned $18 \times 18$ binary metal layer (4\,$\mu$m thick) is embedded between two dielectrics ($\varepsilon_{\text{eff}} = 3.9$) over a ground plane, with 2-port $S$-parameters sampled at 36 frequencies from 30 to 100\,GHz. (b) Two ports (red) on the top/bottom edges carry signal and two AC shorts (blue) on the left/right edges are present in all cavities. Cross-edge and same-edge port arrangements are both included in training, with port pixel locations randomized. (c) NMD pipeline. CavityNet encodes the geometry into cavity tokens, PoleNet maps the tokens to $K$ complex poles $p_k$, and AmpNet (left/right) is queried per port to produce $l_{k,i}$ and $r_{k,j}$. The entry-wise synthesis $Z_{ij}(\omega) = \sum_k l_{k,i} r_{k,j}/(j\omega - p_k)$ is converted to $S$ and supervised.}
  \label{fig:architecture}
\end{figure}

\subsection{Setup}
\label{sec:setup}
\paragraph{Dataset.} Each cavity instance is represented as an $18 \times 18$ binary pixel matrix corresponding to a physical area of $300 \times 300\,\mu\text{m}^2$ (Figure~\ref{fig:architecture}a,b). The interior $16 \times 16$ region encodes the cavity metal pattern. The left and right edges contain a single metal pixel each that serves as an AC short to ground in all cavities, while two ports are placed at the top and/or bottom edges. The two ports may sit on the same edge (Figure~\ref{fig:architecture}b). Port pixel locations along each edge are randomized.

For each instance, we obtain the 2-port $S$-parameter response from full-wave electromagnetic simulation with the commercial simulator EMX, at 36 frequency points spanning $30$ to $100$~GHz in $2$~GHz increments, retaining $\{S_{11}, S_{12}, S_{22}\}$ by reciprocity. Full simulation and dataset construction details follow \citep{karahan2024deepmultiport}. The training set comprises approximately 291{,}000 cavity samples. Test sets for zero-shot port-count extension consist of 3-, 4-, and 5-port configurations on cavity patterns never seen at any port count during training, with $1{,}000$ samples each. Full split sizes are reported in Appendix~\ref{app:dataset}.

\paragraph{Implementation.} We train with AdamW (learning rate $3 \times 10^{-4}$, weight decay $10^{-3}$, batch size 256) for 200 epochs, which takes about 1 hour on one H200 GPU. Loss weights are $\lambda_Z = 0.1$ and $\lambda_g = 0.5$, tuned on the validation set. The total model has approximately 14.5M parameters (CavityNet 4.11M, PoleNet 1.99M, two AmpNet heads 4.20M each). Across 10 random seeds the best validation loss varies with a coefficient of variation of $2.9\%$ (Appendix~\ref{app:seeds}), and removing a random $12.5\%$ or $20.5\%$ of the training data raises the 2-port MAE only gradually (Appendix~\ref{app:pairs}). Layer specifications, hyperparameters, and the full training protocol are detailed in Appendix~\ref{app:training}.

\begin{figure}[!htbp]
  \centering
  \includegraphics[width=0.85\textwidth]{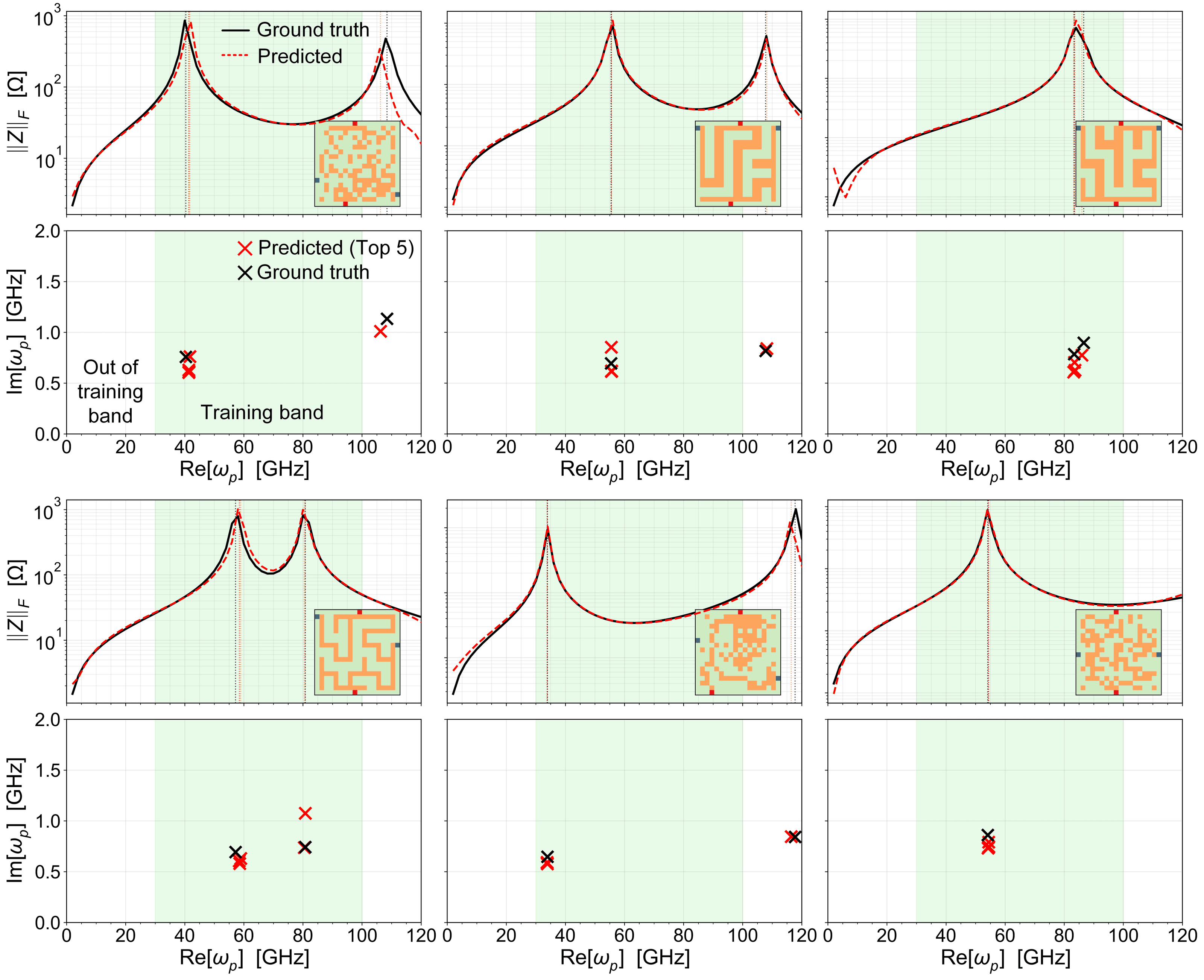}
  \caption{\textbf{Network discovers physical poles without supervision.} Six representative test geometries (the cavity pattern is shown inset in each panel). For each, the top panel shows the Frobenius norm $\|Z(\omega)\|_F$ with ground truth (solid) versus prediction (dashed), and the training band shaded green. The bottom panel shows the top-5 dominant poles in the complex frequency plane, with network poles (red) overlapping AAA-extracted reference poles (black) despite no pole supervision. Agreement holds inside and outside the training band. Full test-set statistics are given in Section~\ref{sec:aaa}.}
  \label{fig:aaa}
\end{figure}

\subsection{Emergent physical pole discovery}
\label{sec:aaa}

To test whether our freely parameterized poles correspond to physically meaningful modes, we compare them against poles extracted by the AAA rational approximation algorithm \citep{nakatsukasa2018aaa} applied independently to the same $S$-parameter data. AAA does not give the exact physical poles, but it is one of the more accurate established methods for extracting them from a sampled response, so agreement with it is evidence of, not proof of, physical correctness (Section~\ref{sec:rlc} tests against exact poles). The reference is stable. Removing one sixth of the frequency samples or varying its convergence threshold moves the poles by less than $0.001\%$ of the band, so the deviations below are attributable to the model.

Figure~\ref{fig:aaa} shows the comparison on six representative test geometries. For each geometry, the top panel plots the Frobenius norm of the impedance matrix, $\|Z(\omega)\|_F = \big(\sum_{i,j} |Z_{ij}(\omega)|^2\big)^{1/2}$, as a single scalar curve summarizing the overall response magnitude across all entries. Resonance peaks in $\|Z(\omega)\|_F$ correspond to poles where physical modes dominate the response. The bottom panel plots the corresponding poles in the complex frequency plane, with the horizontal axis giving the resonance frequency $\omega_k$ and the vertical axis the damping rate $\gamma_k$. Among the $K=32$ network poles we display the five most dominant ones, ranked by the in-band Frobenius contribution of each mode to $\|Z(\omega)\|_F$ (Appendix~\ref{app:topk}). The poles selected this way fall almost exactly on top of the AAA-extracted reference poles, although the network receives no information about pole positions during training, only the system-level $S$-parameter response.

\paragraph{Full test-set statistics.} Matching poles on 3{,}000 test samples (Appendix~\ref{app:polematch}), with distances in the complex frequency plane as a fraction of the 70~GHz band, the median distance from each AAA reference pole to the nearest network pole is $0.504\%$ ($0.35$~GHz), with 75th and 95th percentiles of $1.163\%$ and $3.042\%$, and $70.4\%$ of reference poles lie within $1\%$ and $88.7\%$ within $2\%$. The median resonance-frequency error is $0.48\%$ of the band and the median damping error $0.07\%$. Requiring the match to be one of the five most dominant network poles, as in Figure~\ref{fig:aaa}, $87.4\%$ of reference poles are recovered within $2\%$, and recovery saturates for $K \ge 16$ (Appendix~\ref{app:polematch}).

\begin{figure}[!htbp]
  \centering
  \includegraphics[width=0.82\textwidth]{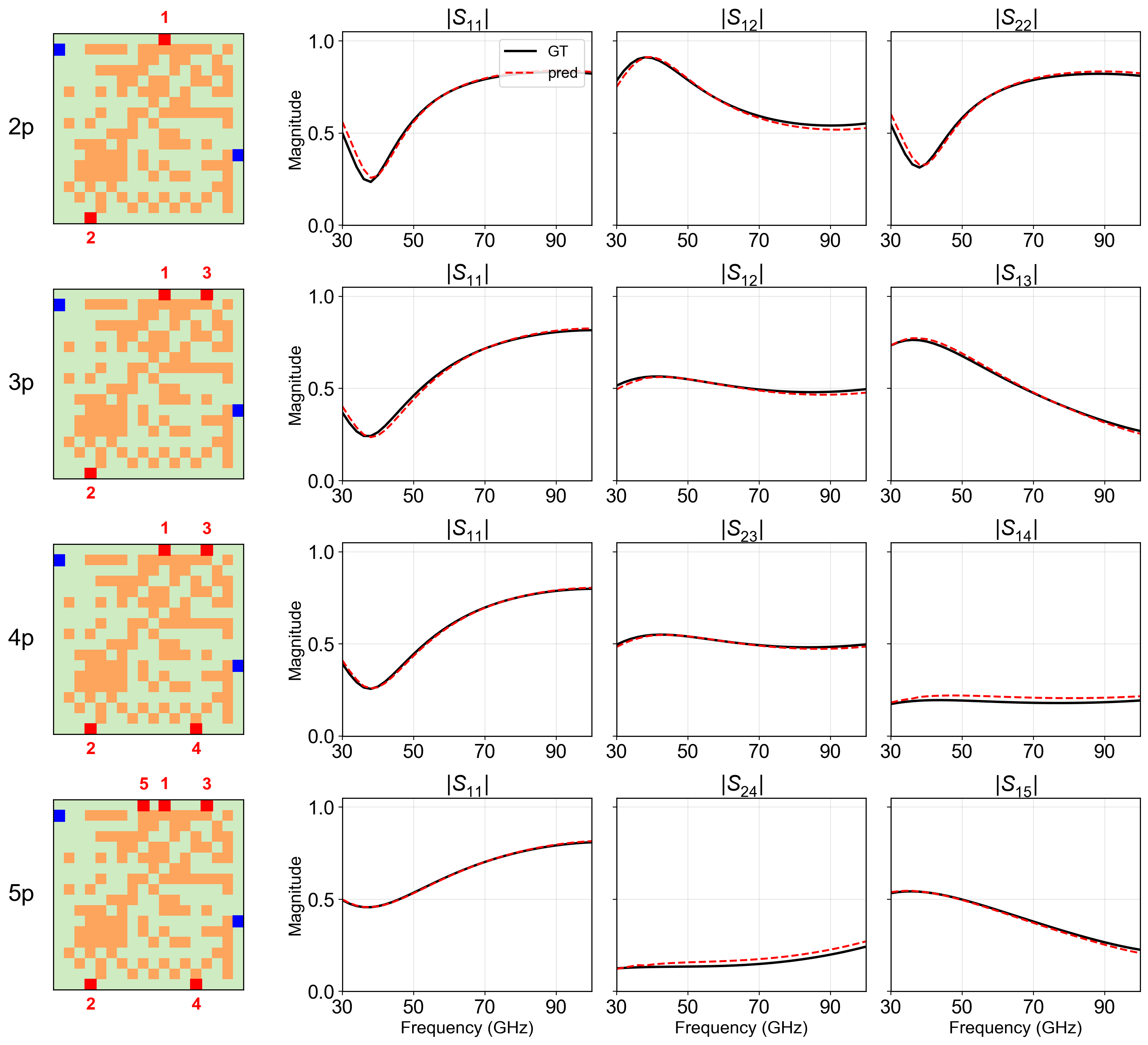}
  \caption{\textbf{Zero-shot port-count extension.} A single model trained only on 2-port data is evaluated on 3-, 4-, and 5-port configurations not seen at any port count during training. Each row corresponds to a port count (2p, 3p, 4p, 5p). The leftmost column shows the cavity with port locations (red, numbered) and AC shorts (blue). Remaining columns show representative $S$-parameter magnitudes versus frequency, with predictions (red, dashed) closely matching ground truth (black, solid).}
  \label{fig:port-extension}
\end{figure}

\subsection{Zero-shot port-count extension}
\label{sec:port-extension}

The network is trained exclusively on 2-port data and tested zero-shot on 3-, 4-, and 5-port configurations. Figure~\ref{fig:port-extension} compares predicted versus ground-truth $S$-parameter responses across all four port counts. On the in-distribution 2-port test set, our model achieves an $S$-parameter MAE of $0.0281 \pm 0.0267$ (mean $\pm$ std). A CNN following \citep{karahan2024deepmultiport} reaches $0.0184$ on this set but maps the whole image to a fixed 2-port output and cannot be evaluated at any other port count. An improved NMD configuration with a full-resolution encoder reaches $0.0112$, below the CNN, and $0.0321$, $0.0354$, and $0.0391$ zero-shot on 3, 4, and 5 ports (Appendix~\ref{app:cnn}). On the held-out port-count test sets ($n=1{,}000$ each), our main model achieves S-parameter MAE of $0.0405 \pm 0.0627$ on 3-port, $0.0433 \pm 0.0602$ on 4-port, and $0.0445 \pm 0.0570$ on 5-port, without ever observing more than two simultaneously active ports during training. As we show in Appendix~\ref{app:error-dist}, the median MAE on these held-out sets is substantially lower than the mean, with the gap driven by a long tail of difficult cavities. For the typical sample, port-count extension adds almost no error to the main model. The submatrix consistency property exploited by our architecture, combined with AmpNet's smooth dependence on port location, yields this generalization. Every port position appears in training, but individual port pairs need not. A model retrained with 54 or 82 port pairs removed entirely from training predicts those pairs as accurately as pairs seen in training (Appendix~\ref{app:pairs}), and with ports on all four edges the same zero-shot assembly extends to 9 ports (Appendix~\ref{app:lr}). Additional zero-shot port-count extension examples are provided in Appendix~\ref{app:more-port-ext}.

\paragraph{Controlled comparison with direct regression.} To isolate the pole-residue form, we train a direct-regression model with the identical CavityNet, per-port query path, data, and protocol, replacing only the NMD head with a multilayer perceptron (MLP) that regresses impedance entries directly (Appendix~\ref{app:directreg}). $N$-port matrices are assembled from pair queries exactly as in NMD. The two models are close on 2-port data, while the median gap grows to $31$--$37\%$ on zero-shot port-count extension (Table~\ref{tab:directreg}). Direct regression retains partial $N$-port accuracy, since each $N$-port entry is close to a 2-port response by submatrix consistency, but only the shared pole set keeps the assembled entries mutually consistent.

\begin{table}[!htbp]
\centering
\small
\begin{tabular}{lcccc}
\toprule
 & \multicolumn{2}{c}{NMD} & \multicolumn{2}{c}{Direct regression} \\
\cmidrule(lr){2-3}\cmidrule(lr){4-5}
Port count & mean & median & mean & median (gap) \\
\midrule
2p & 0.0281 & 0.0221 & 0.0288 & 0.0231 ($+4.4\%$) \\
3p & 0.0405 & 0.0258 & 0.0483 & 0.0353 ($+36.8\%$) \\
4p & 0.0433 & 0.0250 & 0.0500 & 0.0328 ($+31.0\%$) \\
5p & 0.0445 & 0.0245 & 0.0511 & 0.0322 ($+31.4\%$) \\
\bottomrule
\end{tabular}
\caption{NMD versus a direct-regression head sharing the same encoder, port queries, data, and training protocol. $S$-parameter MAE on the same test sets as Table~\ref{tab:error-dist}. The gap is relative to the NMD median.}
\label{tab:directreg}
\end{table}

\subsection{Validation against exactly known poles in RLC networks}
\label{sec:rlc}

To test pole recovery against exact ground truth, and to check that the results are not tied to the pixel representation or the restricted port placement of the RF dataset, we built a second system class with analytic poles, namely random RLC networks with 8 nodes (6 port candidates), whose poles are exactly the eigenvalues of the circuit's state matrix. The circuit graph with its component values is encoded by a small graph network in place of CavityNet, while the NMD head is unchanged and ports are arbitrary circuit nodes. Trained on 48{,}000 circuits observed only through 2-port responses, the model reaches $S$-parameter MAE $0.0178$ on 2-port responses of 2{,}000 held-out circuits and $0.0157$, $0.0141$, and $0.0129$ zero-shot on 3-, 4-, and 5-port responses (Table~\ref{tab:rlc}, Appendix~\ref{app:rlc}), so the error does not grow with the number of ports. Against the 9{,}329 exact in-band poles of the test circuits, $97.9\%$ have a network pole within $1\%$ of the band and $99.8\%$ within $2\%$, with a median resonance-frequency error of $0.15\%$ of the band. Because PoleNet initializes its poles on a uniform grid across the band, an untrained network already places a pole near many true poles. For this untrained network the corresponding rates are $15.7\%$ and $64.9\%$, so the $1\%$ tolerance is the informative one. The benchmark requires no proprietary tools and can be regenerated from the specification in Appendix~\ref{app:rlc}.

\subsection{Why does emergence happen?}

We observe that the dominant in-band network poles coincide with AAA reference poles across the test set (Section~\ref{sec:aaa}) and with the exact poles of the RLC networks (Section~\ref{sec:rlc}), and recovery saturates once $K$ is large enough to cover the region of the complex plane where poles occur (Appendix~\ref{app:polematch}). We conjecture that a band-limited observable tightly constrains its dominant in-band poles, because any pole-residue fit of the response must place poles near the resonances that shape it. The architecture would then narrow the solution space to a neighborhood of the physical decomposition, and observable supervision would select a point within it. This argument does not extend to the additional poles outside the band, which absorb the smooth part of the response and need not be unique.

\subsection{Limitations}
\label{sec:limitations}
We learn $K$ independent poles, more than the number of physical poles extracted by AAA, and the additional poles absorb the frequency-independent contribution of the pole-residue form. How these distant poles continue to support accurate port-count extension to unseen configurations remains an open question. Our framework also targets LTI systems with discrete modal spectra, so continuous-spectrum or strongly nonlinear regimes lie outside its scope. Finally, the pole count $K$ is a fixed hyperparameter rather than learned, with adaptive selection left open.

\section{Conclusion}
\label{sec:conclusion}

We presented a neural framework for modal decomposition that learns physically meaningful pole-residue representations from system-level observables alone, without supervising the modal parameters themselves. By embedding pole-residue structure as an architectural prior with a factorization that automatically makes the loss invariant to modal sign ambiguities, our framework avoids the central pathologies of prior modal-supervised approaches. The learned poles agree with poles extracted by AAA and with the exact poles of RLC networks, and a single 2-port-trained model generalizes zero-shot to port counts unseen during training, removing the need to train a separate model for each port count. Beyond the RF domain, the same architectural prior could potentially extend to other LTI multi-port systems whose responses admit a pole-residue form, such as photonic devices and superconducting quantum chips, which we leave to future work. More broadly, the pole-residue form does not just describe the response of a multi-port LTI system. It carries the system's eigenmodes inside its algebraic structure. When the observable already has this kind of structure, putting it into the architecture is enough. The network learns the modes from observables, with no need for field training or PDE-residual losses.

This points to a more general reading of the same construction. Nothing in the architecture requires the query to be a physical port. The coupling $l_{k,i}$ is the modal field evaluated at a location, and the port index is merely a discrete sampling of it. Whenever the observable is a Green's function,
\begin{equation}
G(\mathbf{x}, \mathbf{x}', \omega) = \sum_{k} \frac{\phi_k(\mathbf{x})\,\phi_k(\mathbf{x}')}{j\omega - p_k},
\end{equation}
that is, a resolvent sampled at pairs of spatial points rather than at ports, the same factorization applies verbatim, with the port query replaced by a continuous coordinate and the residue product by the modal field profiles $\phi_k$. The framework may therefore not be confined to port-level observables but could extend to spatially resolved ones, yielding a mesh-free modal representation of the field. We leave this to future work.

\begin{ack}
This work was supported by the Army Research Office under Award No.\ W911NF2410111 (P.I.\ Kaushik Sengupta). The authors have filed a United States provisional patent application on 5 September 2026, \#64/148,880, entitled ``AI-ENABLED METHOD AND SYSTEM FOR PREDICTING MULTI-PORT DEVICE RESPONSES USING PHYSICS-EMBEDDED MODAL DECOMPOSITION.''
\end{ack}

\bibliographystyle{unsrtnat}
\bibliography{reference}


\appendix

\section{Dataset details}
\label{app:dataset}

The cavity dataset is generated by sampling binary metal patterns over the interior $16 \times 16$ region of the $18 \times 18$ pixel array, together with port pixel locations along the top and bottom edges. Of the 291{,}435 training samples, 226{,}603 ($77.8\%$) use i.i.d.\ Bernoulli$(0.5)$ pixels and 64{,}832 ($22.2\%$) use maze patterns generated with MATLAB's \texttt{mapMaze}. We mix the two because i.i.d.\ sampling almost never produces long connected metal paths, while practical RF passives are built from connected transmission-line structures. Port and AC-short pixel locations are sampled uniformly along their edges. The full $18 \times 18$ array represents a physical area of $300 \times 300\,\mu\text{m}^2$, simulated with the commercial electromagnetic simulator EMX. Although the simulation uses a 4-port representation in which the AC-shorted left and right edges are formally treated as ports, only the top and bottom edges carry signal and are used as observation ports. The 2-port S-parameter response is stored as $\{S_{11}, S_{12}, S_{22}\}$ at 36 frequency points, exploiting reciprocity ($S_{12} = S_{21}$). Full simulation, sampling, and post-processing details follow \citep{karahan2024deepmultiport}.

\paragraph{Simulator and stack.} EMX is a full-wave method-of-moments solver for planar multilayer structures, which solves Maxwell's equations in integral form on the metal surfaces. The stack (Figure~\ref{fig:architecture}a) consists of a patterned copper layer ($4\,\mu$m thick) between two dielectric layers ($8.3$ and $10.78\,\mu$m thick) over a copper ground plane. The exact layer definitions belong to a proprietary foundry process design kit, which is why the RF dataset cannot be released. The pixel pitch is $16.67\,\mu$m while each metal pixel measures $20.83 \times 20.83\,\mu\text{m}^2$, so adjacent pixels, including diagonal neighbors, overlap by $2.08\,\mu$m per side and are electrically connected. For the same reason, two ports on adjacent pixels of one edge would be electrically the same node and are excluded, leaving 466 realizable 2-port pairs, namely $16 \times 16 = 256$ top--bottom pairs plus $105$ each of top--top and bottom--bottom pairs.

\paragraph{Port reduction.} The left and right edge pixels are formal ports in EMX, so a 2-port geometry is simulated as a 4-port and the two edge ports are short-circuited in post-processing to obtain the 2-port response. The 3-, 4-, and 5-port geometries are likewise simulated as 5-, 6-, and 7-port structures and reduced. Each port count is a separate full-wave simulation with the additional port metal physically present, so an $N$-port ground truth is not extracted from any other simulation.

The 2-port dataset is split into 291{,}435 train, 26{,}429 validation, 
and 26{,}430 test samples. Test sets for zero-shot port-count extension consist of 3-, 4-, and 5-port configurations on cavity patterns not seen at any port count during training, with $1000$ samples each.

Each sample took about 10 seconds to simulate, and generating the whole dataset took about 1.5 days with 50 simulations running in parallel on a cluster of Intel Ice Lake CPUs.

\section{Training details}
\label{app:training}

\subsection{CavityNet specification}

The input to CavityNet is the cavity pattern $g \in \{0,1\}^{1 \times H \times W}$, a binary image in which each pixel is either metal (1) or empty (0). The image has $H = 16$ rows and $W = 18$ columns. The two rows that hold the ports are left out, because port locations are given separately to AmpNet.

CavityNet is a convolutional neural network made of four blocks of convolutional layers. Each convolutional layer slides a learned $3 \times 3$ filter over the image. Its output is then normalized with batch normalization, which rescales each channel using statistics computed over the training batch, and passed through the GELU activation function, a smooth variant of the rectified linear unit. After the first and second blocks, a $2 \times 2$ max-pooling step halves the height and width of the feature map by keeping the largest value in each $2 \times 2$ window. The number of feature channels grows from block to block as follows.
\begin{itemize}
\item Block 1: two convolutional layers with 64 channels, followed by max-pooling ($16 \times 18 \to 8 \times 9$).
\item Block 2: three convolutional layers with 128 channels, followed by max-pooling ($8 \times 9 \to 4 \times 5$).
\item Block 3: three convolutional layers with 256 channels.
\item Block 4: two convolutional layers with $D_t = 384$ channels.
\end{itemize}

The output of block 4 is a $4 \times 5$ grid with a $D_t$-dimensional feature vector at each grid position. We treat each of these vectors as a token that summarizes one region of the cavity, which gives $n = 20$ tokens. Each token is normalized with layer normalization, which shifts and rescales the vector to zero mean and unit variance and then applies a learned scale and offset. A learned position vector is then added to each token so that the later modules know which region of the cavity the token describes. Collecting the position vectors in $E \in \mathbb{R}^{n \times D_t}$, the cavity tokens are
\begin{equation}
T = \text{LayerNorm}\big(\text{Flatten}(\phi_{\text{cnn}}(g))\big) + E \;\in\; \mathbb{R}^{n \times D_t},
\end{equation}
where $\phi_{\text{cnn}}$ denotes the four convolutional blocks and Flatten arranges the $4 \times 5$ grid into a list of 20 tokens. Two kinds of noise are applied during training only, to keep the model from overfitting. Gaussian noise with standard deviation $\sigma_{\text{pat}} = 0.05$ is added to the input pattern, and token dropout sets each token to zero with probability $p_{\text{tok}} = 0.10$.

\subsection{PoleNet specification and pole convention}
\label{app:polenet}

PoleNet turns the 20 cavity tokens into a fixed number $K = 32$ of poles. It does so with $K$ learned query vectors $Q \in \mathbb{R}^{K \times D_t}$, one for each pole. The query vectors are trainable parameters that do not depend on the input cavity. At initialization they are spread out along one coordinate, so that different queries start from different values and do not all converge to the same pole.

Each query collects information from the cavity tokens through attention. The query computes a similarity score with every cavity token, the scores are converted into weights that sum to one, and the query receives the weighted average of the tokens. We run $h = 4$ such attention operations in parallel (``heads''), each with its own learned projections, and combine their outputs. This lets one query draw on several regions of the cavity at once. After attention, each query passes through a small two-layer fully connected network, applied to each query separately, with a hidden layer of width $512$. One attention step followed by one fully connected network forms a block, and PoleNet stacks $L_{\text{pole}} = 2$ blocks. Each block updates the query vectors $x$ as
\begin{equation}
x \leftarrow \text{LayerNorm}\big(x + \text{Attention}(x, T)\big), \qquad x \leftarrow x + \text{MLP}(x),
\end{equation}
where $\text{Attention}(x, T)$ is the attention from the queries $x$ to the cavity tokens $T$ described above, and $\text{MLP}$ is the two-layer fully connected network. Each step adds its output to its input, so that each block only has to learn a correction to the current query vectors.

A final linear layer maps each of the $K$ query vectors to two real numbers $(u_k, v_k)$, which define the parameters of Equation~\ref{eq:pole-param} as
\begin{equation}
\gamma_k = \text{softplus}(u_k), \qquad \omega_k = v_k, \qquad p_k = -\gamma_k + j\,\omega_k.
\end{equation}
The softplus function, $\text{softplus}(u) = \log(1 + e^{u})$, is positive for every input. It therefore guarantees $\gamma_k > 0$, which places every pole in the left half of the $s$-plane. In Figure~\ref{fig:aaa} we display poles in the complex frequency plane through the substitution $\omega_p \equiv -j\, p_k = \omega_k + j\,\gamma_k$. The horizontal axis $\mathrm{Re}\,\omega_p$ is then the resonance frequency $\omega_k$, and the vertical axis $\mathrm{Im}\,\omega_p$ is the damping rate $\gamma_k$.

\subsection{AmpNet specification}

AmpNet is queried with one port at a time. A port is identified by its row $r \in \{0, 17\}$ and its column $c \in \{0, \dots, 17\}$ on the $18 \times 18$ grid, and AmpNet first converts this location into a vector. The row and the column are converted in different ways.

The row index takes only two values in our setup, $r = 0$ for the top edge and $r = 17$ for the bottom edge. A sinusoidal encoding of the normalized value $r/17$ would return nearly identical vectors for these two rows, leaving the network almost no signal to tell the edges apart. We therefore use a learned lookup table for the row, in which each row index has its own trainable vector of length $d_{\text{row}} = 64$,
\begin{equation}
\mathrm{row\_emb}: \{0, 1, \dots, 17\} \to \mathbb{R}^{d_{\text{row}}}, \qquad d_{\text{row}} = 64,
\end{equation}
with entries initialized from a Gaussian distribution with standard deviation $0.05$. The table has 18 entries even though only $r = 0$ and $r = 17$ are used in our experiments, so that configurations with ports on intermediate rows can be added later without changing the model.

The column index takes 18 distinct values along the edge, and for it a sinusoidal encoding works well. The encoding represents the column by sines and cosines at $L_{\text{port}} = 12$ frequencies,
\begin{equation}
\mathrm{col\_enc}(c) = \big[\sin(2^l \pi\, c/17), \cos(2^l \pi\, c/17)\big]_{l=0,\dots,L_{\text{port}}-1} \in \mathbb{R}^{2 L_{\text{port}}}.
\end{equation}
The low-frequency terms separate columns that are far apart, and the high-frequency terms separate neighboring columns. The row vector and the column vector are concatenated and multiplied by a learned matrix $W_{\text{in}}$, which gives a single query vector for the port,
\begin{equation}
q = W_{\text{in}}\,\big[\mathrm{row\_emb}(r) \,\|\, \mathrm{col\_enc}(c)\big] \;\in\; \mathbb{R}^{D_t},
\end{equation}
where $\|$ denotes concatenation.

The port query then collects information from the cavity tokens $T$ through $L_{\text{amp}} = 3$ blocks of the same form as in PoleNet, each made of attention with $h = 4$ heads followed by a two-layer fully connected network, here with a hidden layer of width $1024$. During training, dropout randomly sets $15\%$ of the intermediate values to zero. The output of the last block thus describes the port as seen from inside this particular cavity. Two separate linear layers then read out $2K$ real numbers each, a magnitude and a phase for every pole. One layer gives the left coupling $l_{k,i}$, used when the port is the row index $i$ of $Z_{ij}$, and the other gives the right coupling $r_{k,j}$, used when the port is the column index $j$. The magnitudes pass through a softplus so that they are never negative, and the phases are left unconstrained.

\subsection{Loss, gauge term, and optimization}
\label{app:gauge}

The total training loss is
\begin{equation}
\mathcal{L} = \mathcal{L}_S + \lambda_Z\, \mathcal{L}_{Z,\log} + \lambda_g\, \mathcal{L}_{\text{gauge}}.
\end{equation}
$\mathcal{L}_S$ is the mean absolute error between the predicted and ground-truth scattering matrices, averaged over the real and imaginary parts of all matrix entries and over all frequencies. $\mathcal{L}_{Z,\log}$ compares the impedance matrices on a logarithmic scale, as the mean absolute difference between $\log(|Z|+\varepsilon)$ of the prediction and of the ground truth. The logarithm keeps the very large impedance values near resonance from dominating the loss, and the small constant $\varepsilon = 10^{-2}$ keeps the logarithm finite when $|Z|$ is close to zero. The loss weights are $\lambda_Z = 0.1$ and $\lambda_g = 0.5$.

\paragraph{Gauge term in detail.} Section~\ref{sec:method} introduced the gauge-fixing strategy and the per-mode penalty (Equation~\ref{eq:gauge-fix}). Here we give the full form. Each training sample queries two ports. Let $\mathcal{P}_{\text{anchor}}$ be the set of those ports that lie on the anchor edge, the bottom edge $r_{\mathrm{anchor}} = 17$. The gauge loss of a sample is the per-mode penalty averaged over both AmpNet outputs (left and right), over the $K$ modes, and over the ports in $\mathcal{P}_{\text{anchor}}$,
\begin{equation}
\mathcal{L}_{\text{gauge}} = \frac{1}{2 K\, |\mathcal{P}_{\text{anchor}}|} \sum_{p \in \mathcal{P}_{\text{anchor}}} \sum_{k=1}^K \big[\ell_{\text{gauge}}(l_{k,p}) + \ell_{\text{gauge}}(r_{k,p})\big],
\end{equation}
with $\varepsilon_{\text{anchor}} = 10^{-6}$. When neither port lies on the bottom edge, $\mathcal{P}_{\text{anchor}}$ is empty and we set $\mathcal{L}_{\text{gauge}} = 0$. In other words, a sample with both ports on the top edge contributes no gauge loss, a sample with one port on each edge anchors only the bottom-edge port, and a sample with both ports on the bottom edge anchors both ports. In every case the anchored port is constrained through both the left and the right coupling.

\paragraph{Optimization.} We train with the AdamW optimizer, a variant of Adam that applies weight decay (a pull of the weights toward zero) separately from the gradient update. We use learning rate $3 \times 10^{-4}$, weight decay $10^{-3}$, and batch size 256, for 200 epochs. The model has about 14.5M parameters in total, with $4.11$M in CavityNet, $1.99$M in PoleNet, and $4.20$M in each of the left and right AmpNet outputs.

\begin{figure}[!htbp]
  \centering
  \includegraphics[width=0.75\textwidth]{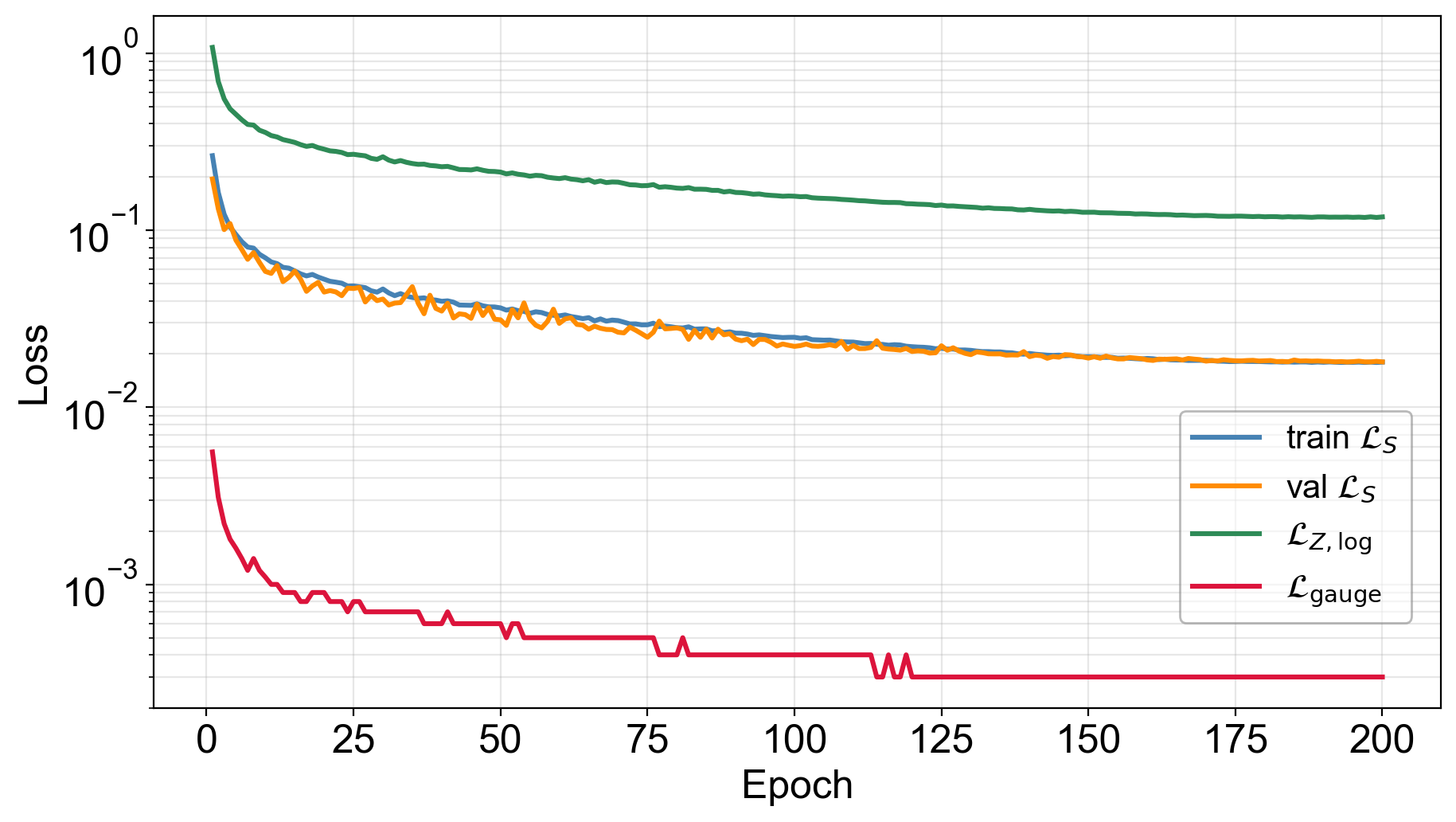}
  \caption{\textbf{Training dynamics over 200 epochs.} Train and validation 
  $\mathcal{L}_S$ (S-parameter MAE), $\mathcal{L}_{Z,\log}$ (log-impedance error), 
  and $\mathcal{L}_{\text{gauge}}$. Train and validation $\mathcal{L}_S$ track closely with no 
  overfitting. $\mathcal{L}_{\text{gauge}}$ saturates near $3 \times 10^{-4}$, 
  indicating near-complete gauge fixing.}
  \label{fig:learning-curve}
\end{figure}

\paragraph{Training dynamics.} Figure~\ref{fig:learning-curve} shows the 
training and validation $\mathcal{L}_S$, the log-impedance loss $\mathcal{L}_{Z,\log}$, 
and the gauge-fixing loss $\mathcal{L}_{\text{gauge}}$ over 200 epochs. 
The training and validation $\mathcal{L}_S$ curves track each other closely 
throughout, indicating no overfitting on the in-distribution 2-port data. 
The gauge-fixing loss $\mathcal{L}_{\text{gauge}}$ converges to roughly 
$3 \times 10^{-4}$, indicating near-complete gauge fixing on the anchor edge.

\subsection{Hyperparameter summary}
The hyperparameters used in this paper are stated in Table~\ref{tab:hyperparams}.

\begin{table}[!htbp]
\centering
\small
\begin{tabular}{llc}
\toprule
Component & Parameter & Value \\
\midrule
CavityNet & token dimension $D_t$ & 384 \\
          & number of tokens $n$ & $4 \times 5 = 20$ \\
          & pattern noise $\sigma_{\text{pat}}$ / token dropout $p_{\text{tok}}$ & 0.05 / 0.10 \\
\midrule
PoleNet   & number of poles $K$ & 32 \\
          & blocks / attention heads / hidden width & 2 / 4 / 512 \\
\midrule
Left-AmpNet  & row embedding dim $d_{\text{row}}$ & 64  \\
             & column encoding frequencies $L_{\text{port}}$ & 12 \\
             & blocks / attention heads / hidden width & 3 / 4 / 1024 \\
             & dropout & 0.15 \\
\midrule
Right-AmpNet & (identical architecture to Left-AmpNet) & same \\
\midrule
Loss      & $\lambda_Z$ & 0.1 \\
          & $\lambda_g$ (anchor weight) & 0.5 \\
          & log-$Z$ stabilizer $\varepsilon$ & $10^{-2}$ \\
          & anchor stabilizer $\varepsilon_{\text{anchor}}$ & $10^{-6}$ \\
          & anchor row $r_{\mathrm{anchor}}$ & 17 \\
\midrule
Optimization & optimizer & AdamW \\
             & learning rate / weight decay & $3 \times 10^{-4}$ / $10^{-3}$ \\
             & batch size / epochs & 256 / 200 \\
\midrule
Parameter breakdown & CavityNet & $4.11$M \\
                    & PoleNet & $1.99$M \\
                    & Left-AmpNet & $4.20$M \\
                    & Right-AmpNet & $4.20$M \\
\midrule
Total parameters & & $\sim$14.5M \\
\bottomrule
\end{tabular}
\caption{Hyperparameters for the neural modal decomposition model. The two AmpNet heads share the same architecture but have independent weights, producing left coupling $l_{k,i}$ and right coupling $r_{k,j}$ respectively. No penalty pushing $l_{k,i}$ toward $r_{k,i}$ (which would enforce reciprocity) is used, so that the framework remains applicable to non-reciprocal systems.}
\label{tab:hyperparams}
\end{table}

\section{Top-5 dominance criterion for pole comparison}
\label{app:topk}

Figure~\ref{fig:aaa} compares network poles against AAA poles using $\|Z(\omega)\|_F$, the Frobenius norm of the impedance matrix, as a scalar summary of the response. For each mode $k$ we therefore want a dominance score that captures its in-band contribution to $\|Z(\omega)\|_F$, so that the top-ranked modes are precisely those that shape the curves shown in the figure.

For a sample with two queried ports $i$ and $j$, mode $k$ contributes to the $2 \times 2$ impedance submatrix through the rank-1 residue matrix
\begin{equation}
M_k = \mathbf{l}_k\, \mathbf{r}_k^T = \begin{bmatrix} l_{k,i} \\ l_{k,j} \end{bmatrix} \begin{bmatrix} r_{k,i} & r_{k,j} \end{bmatrix},
\qquad
Z(\omega) = \sum_{k=1}^K \frac{M_k}{j\omega - p_k}.
\end{equation}
The in-band $L^2$ contribution of mode $k$ to $\|Z(\omega)\|_F$ on the training band is then
\begin{equation}
\sqrt{\int_{\omega \in \text{band}} \left\| \frac{M_k}{j\omega - p_k} \right\|_F^2 \, d\omega} \;=\; \|M_k\|_F \cdot \sqrt{\int_{\omega \in \text{band}} \frac{1}{|j\omega - p_k|^2} \, d\omega},
\end{equation}
a product of the size of the residue matrix and a gain factor that is large when the pole lies close to the band.

The rank-1 identity $\|M_k\|_F^2 = \|\mathbf{l}_k\|^2 \cdot \|\mathbf{r}_k\|^2$ lets us evaluate the residue factor directly from the network's left and right couplings, and we replace the integral with a discrete sum over the $N_f = 36$ training frequencies $\{\omega_n\}_{n=1}^{N_f}$ uniformly spaced in $[30, 100]$~GHz. The dominance score is
\begin{equation}
\mathrm{dom}_k = \underbrace{\sqrt{\big(|l_{k,i}|^2 + |l_{k,j}|^2\big)\big(|r_{k,i}|^2 + |r_{k,j}|^2\big)}}_{\|M_k\|_F} \cdot \underbrace{\sqrt{\sum_{n=1}^{N_f} \frac{1}{|j\omega_n - p_k|^2}}}_{\text{in-band kernel gain}}.
\end{equation}
We rank modes by $\mathrm{dom}_k$ and select the top $N_{\text{TOP}} = 5$.

Either factor alone is misleading. The Frobenius norm alone selects modes with large amplitude that may sit far outside the training band and contribute little to $\|Z(\omega)\|_F$ on it. The kernel gain alone selects well-placed poles that may have negligible residue, including spurious modes. Their product is the in-band Frobenius contribution we want.

\section{Quantitative pole matching}
\label{app:polematch}

\paragraph{Reference poles.} For each test sample we apply AAA \citep{nakatsukasa2018aaa} to the full $2 \times 2$ ground-truth impedance $Z(\omega)$ at the 36 training frequencies (tolerance $10^{-3}$, maximum degree 16, removal of spurious pole-zero pairs known as Froissart doublets at $10^{-3}$, retaining only stable poles). A reference pole is kept if it reappears when each of six contiguous frequency blocks is removed in turn (at least 5 of 6 refits) and if its quality factor $Q = \omega_k / (2\gamma_k)$ is at least 2. The same $Q \ge 2$ filter is applied to network poles, which otherwise admit very broad poles that absorb a constant offset. On 3{,}000 test samples this yields 2{,}358 in-band reference poles. Varying the AAA tolerance and the Froissart threshold by a factor of two, or removing one sixth of the frequency samples, moves the retained poles by less than $0.001\%$ of the band.

\paragraph{Matching criterion.} Several network poles can sit near the same physical pole, so a one-to-one assignment is not appropriate. We report the distance from each reference pole to the nearest network pole, and the recovery rate, defined as the fraction of reference poles for which one of the five most dominant network poles (Appendix~\ref{app:topk}) lies within a given tolerance. A tolerance of $2\%$ of the band ($1.4$~GHz) is about 0.9 times the typical linewidth of a mode. As a chance baseline, matching each sample against the predicted poles of a randomly chosen other sample gives a recovery rate of $4$--$6\%$ at $2\%$.

\paragraph{Sensitivity to the pole count $K$.} Table~\ref{tab:pole-recovery-K} reports the recovery rate for models trained with $K \in \{8, 16, 32, 64\}$.

\begin{table}[!htbp]
\centering
\small
\begin{tabular}{lcc}
\toprule
$K$ & recovered within $1\%$ & recovered within $2\%$ \\
\midrule
8  & $37.7\%$ & $67.5\%$ \\
16 & $60.3\%$ & $85.6\%$ \\
\textbf{32 (main result)} & $\mathbf{69.1\%}$ & $\mathbf{87.4\%}$ \\
64 & $68.9\%$ & $89.3\%$ \\
\bottomrule
\end{tabular}
\caption{Pole recovery rate against AAA reference poles (top-5 dominant network poles, 3{,}000 test samples) as a function of the pole count $K$.}
\label{tab:pole-recovery-K}
\end{table}

\paragraph{Choosing $K$.} $K$ should not be read as the number of physical resonances of a given system. Pole locations vary strongly from geometry to geometry, so the $K$ predicted poles must cover the whole region of the complex frequency plane in which poles occur across the geometry family, with each sample activating only a subset of them as its dominant resonances. Undersizing $K$ immediately lowers recovery ($K=8$), while beyond a sufficient value recovery plateaus and oversizing is harmless (doubling from 32 to 64 changes recovery at $2\%$ by about two points). A practical procedure for a new system is therefore to sweep $K$ over a few values and select the smallest value at which recovery and validation error saturate.

\section{Error distribution across port counts}
\label{app:error-dist}

Section~\ref{sec:port-extension} summarizes zero-shot generalization through the mean $S$-parameter MAE on each port count. Here we report the shape of the error distribution in more detail. Table~\ref{tab:error-dist} reports the median, the 25th and 75th percentiles (p25, p75), and the 95th percentile (p95) on the 2-port in-distribution test set ($n=26{,}430$) and on the held-out 3-, 4-, and 5-port test sets ($n=1{,}000$ each).

\begin{table}[!htbp]
\centering
\small
\setlength{\tabcolsep}{4pt}
\begin{tabular}{lccccccccc}
\toprule
Port & Test set size & mean & median & std & p25 & p75 & p95  \\
\midrule
2p & 26{,}430 & 0.0281 & 0.0221 & 0.0267 & 0.0152 & 0.0328 & 0.0612 \\
3p & 1{,}000  & 0.0405 & 0.0258 & 0.0627 & 0.0184 & 0.0362 & 0.1075 \\
4p & 1{,}000  & 0.0433 & 0.0250 & 0.0602 & 0.0184 & 0.0376 & 0.1794 \\
5p & 1{,}000  & 0.0445 & 0.0245 & 0.0570 & 0.0183 & 0.0363 & 0.1925\\
\bottomrule
\end{tabular}
\caption{Distribution of per-sample $S$-parameter MAE across port counts. The mean reported in the main text is included for reference. Across all held-out port counts the median sits substantially below the mean, indicating that the generalization error is concentrated in a small upper tail rather than spread uniformly across the test set.}
\label{tab:error-dist}
\end{table}

The mean is raised by a long tail of difficult samples, while the typical sample generalizes with almost no additional error. On every held-out port count the median MAE lies well below the mean ($0.0258$ vs.\ $0.0405$ for 3p, $0.0250$ vs.\ $0.0433$ for 4p, and $0.0245$ vs.\ $0.0445$ for 5p), and the interquartile range stays tightly clustered around it. In particular, p25 is nearly identical across the three held-out port counts ($0.0184$, $0.0184$, $0.0183$) and only slightly above its 2-port value ($0.0152$), and p75 also remains close to its 2-port value. For the easier half of the test distribution, port-count extension therefore adds little error. 

The inflation of the mean at higher port counts is driven almost entirely by the upper tail. p95 grows from $0.0612$ at 2p to $0.1925$ at 5p. A small number of difficult cavities, possibly those whose poles are wrongly predicted, account for most of the gap between the in-distribution and held-out means. The standard deviation reported alongside the mean in Section~\ref{sec:port-extension} reflects this tail rather than dispersion of the typical sample.

\section{Additional port-count extension examples}
\label{app:more-port-ext}

\begin{figure}[!htbp]
  \centering
  \includegraphics[width=0.95\textwidth]{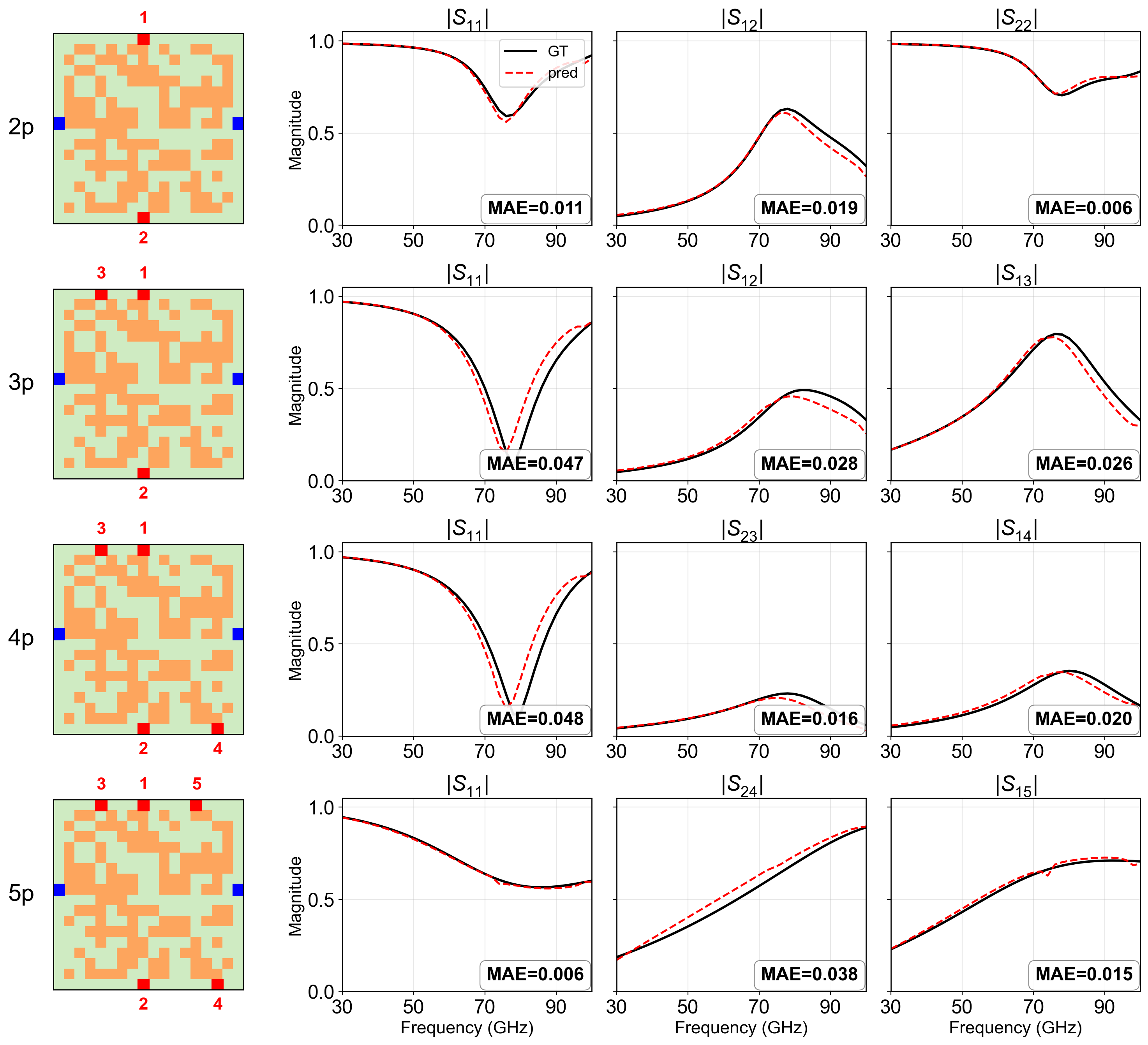}
  \caption{\textbf{Zero-shot port-count extension, Additional example 1.}}
  \label{fig:port-ext-random}
\end{figure}

\begin{figure}[!htbp]
  \centering
  \includegraphics[width=0.95\textwidth]{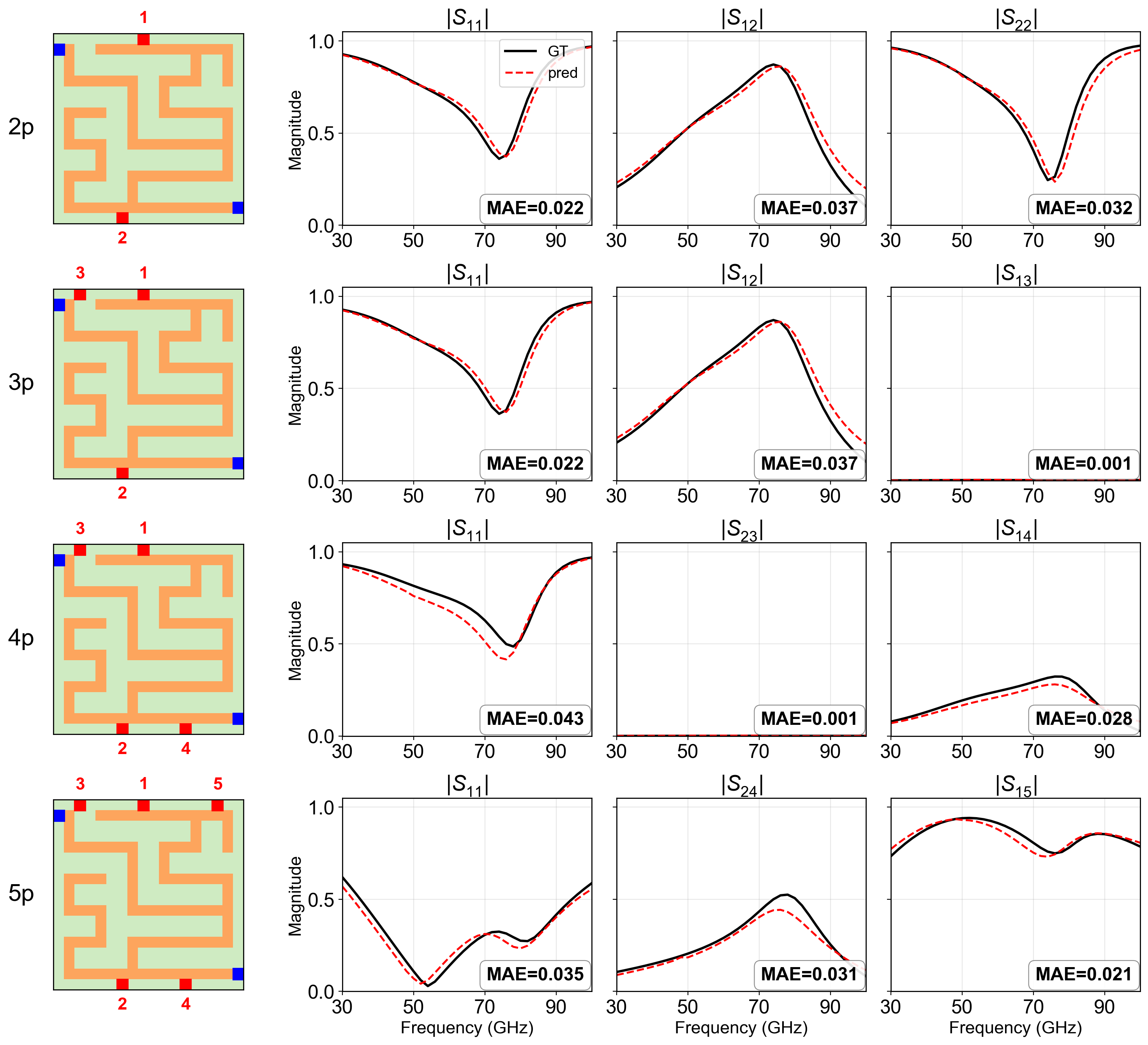}
  \caption{\textbf{Zero-shot port-count extension, Additional example 2.}}
  \label{fig:port-ext-maze}
\end{figure}

Figure~\ref{fig:port-extension} in the main text shows zero-shot port-count extension on a single representative geometry. Here we provide two additional examples in Figures~\ref{fig:port-ext-random} and~\ref{fig:port-ext-maze}. As before, the model is trained exclusively on 2-port data and queried at 3, 4, and 5 ports without any further adaptation.

\section{Ablation studies}
\label{app:ablation}

We ablate two key design choices, the number of poles $K$ and the gauge-fixing term. All other hyperparameters and the training protocol follow Appendix~\ref{app:training}. Table~\ref{tab:ablation} reports the $S$-parameter 
MAE on the in-distribution 2-port test set ($n=26{,}430$) and on the held-out 
3-, 4-, and 5-port test sets used for zero-shot port-count extension ($n=1000$ each).

\begin{table}[!htbp]
\centering
\small
\begin{tabular}{lcccccc}
\toprule
Model & $K$ & $\lambda_g$ & 2p & 3p & 4p & 5p \\
\midrule
K=8                                & 8   & 0.5 & 0.0752 & 0.0995 & 0.0944 & 0.0887 \\
K=16                               & 16  & 0.5 & 0.0306 & 0.0533 & 0.0533 & 0.0569 \\
\textbf{K=32 with gauge fix (main result)} & 32  & 0.5 & \textbf{0.0281} & \textbf{0.0405} & \textbf{0.0433} & \textbf{0.0445} \\
K=32 without gauge fix                 & 32  & 0.0 & 0.0333 & 0.0439 & 0.0438 & \textbf{0.0453} \\
K=64                               & 64  & 0.5 & \textbf{0.0279} & 0.0423 & \textbf{0.0431} & 0.0458 \\
\bottomrule
\end{tabular}
\caption{Ablation on pole count $K$ and gauge-fixing weight $\lambda_g$. 
S-parameter MAE on the 2-port in-distribution test set and on held-out 3-, 4-, 
and 5-port test sets (zero-shot port-count extension). The default configuration 
used throughout the main paper is K=32 with gauge fix.}
\label{tab:ablation}
\end{table}

\paragraph{Pole count $K$.} With $K=8$ the model is severely underparameterized 
and fails on every port count (2p MAE $0.0752$). Increasing to $K=16$ already 
recovers competitive in-band accuracy ($0.0306$) but still degrades on the 
held-out port counts ($0.0533$--$0.0569$). At $K=32$ both in-band accuracy 
and port-count generalization saturate, and $K=64$ yields no further 
improvement ($0.0279$ vs.\ $0.0281$ for 2p and $0.0458$ vs.\ $0.0445$ for 5p). We adopt 
$K=32$ as the default. As discussed in Section~\ref{sec:limitations}, this is 
larger than the number of in-band physical poles extracted by AAA. The extra 
poles absorb the frequency-independent contribution of the pole-residue form 
and place themselves outside the training band.

\paragraph{Gauge-fixing weight $\lambda_g$.} Comparing K=32 with gauge 
($\lambda_g = 0.5$) against K=32 without gauge ($\lambda_g = 0$) isolates the 
contribution of the anchor-gauge term. The two variants differ noticeably on 
in-band 2-port reconstruction ($0.0281$ vs.\ $0.0333$, a roughly $16\%$ 
relative improvement), but are much closer on zero-shot 
port-count extension to 3-, 4-, and 5-port configurations ($0.0405$ vs.\ 
$0.0439$, $0.0433$ vs.\ $0.0438$, $0.0445$ vs.\ $0.0453$). The gauge-fixing 
term is therefore not essential for the port-count generalization capability 
that is the core contribution of this work, but it provides a modest 
improvement in in-band reconstruction fidelity. We retain it as a lightweight 
regularizer in the default configuration.

\section{Direct-regression baseline}
\label{app:directreg}

The direct-regression model of Table~\ref{tab:directreg} keeps CavityNet and the per-port cross-attention query path of AmpNet (three blocks, as in Appendix~\ref{app:training}) but uses a single shared query trunk. For a port pair $(i, j)$ with query features $x_i, x_j$, the head receives the symmetric combination $[x_i + x_j,\; x_i \odot x_j]$, where $\odot$ is the elementwise product, which builds reciprocity into the output, and an MLP (three hidden layers of width 1024) regresses 72 values, the real and imaginary parts of $Z_{ij}$ at the 36 frequencies. There are no poles or residues. It is trained with $\mathcal{L}_S + 0.1\,\mathcal{L}_{Z,\log}$ (the gauge term has no counterpart) on the same data, augmentation, and 200-epoch protocol, and has 11.25M parameters against 14.5M for NMD. Its best validation $\mathcal{L}_S$ is $0.01835$, against $0.01804$ for NMD, so in-band fitting capacity is essentially matched. $N$-port matrices are assembled entry by entry from $N(N+1)/2$ pair queries.

\section{CNN baseline and an improved configuration}
\label{app:cnn}

The CNN baseline follows \citep{karahan2024deepmultiport}. It is a residual network with spatial attention after every convolution (four stages of five residual blocks) followed by fully connected layers of width 4096, and maps the $18 \times 18$ pattern directly to the real and imaginary parts of $S_{11}$, $S_{12}$, and $S_{22}$ at the 36 frequencies (70.6M parameters). It was trained on the same 2-port training data together with an additional class of transmission-line patterns, and is evaluated on the same 2-port test set as Table~\ref{tab:error-dist} with the same per-sample MAE over the full $2 \times 2$ matrix. Because its output has a fixed 2-port layout, it cannot be evaluated at other port counts.

The improved NMD configuration keeps the pole-residue head, the port queries, and supervision by 2-port responses only, and changes the following. CavityNet omits the pooling stages, so each position of the $16 \times 18$ grid gives one token (288 tokens instead of 20). The number of poles is $K = 64$. Each AmpNet query attends to the pole tokens of PoleNet in addition to the cavity tokens, so the predicted couplings can depend on the predicted modes. An auxiliary term pushes apart dominant network poles whose residues are nearly collinear, which discourages several poles from describing the same mode. Regularization is lighter (dropout 0.05, $\sigma_{\text{pat}} = 0.02$, $p_{\text{tok}} = 0.05$), and training runs for 600 epochs with a cosine learning-rate schedule. The model has 15.0M parameters, and results are from a single seed.

Table~\ref{tab:cnn} shows that the improved configuration is more accurate than the CNN on 2-port data, with a mean MAE $1.6\times$ and a median MAE $2.2\times$ lower, and also improves on the main model at every port count.

\begin{table}[!htbp]
\centering
\footnotesize
\setlength{\tabcolsep}{4pt}
\begin{tabular}{lcccccc}
\toprule
Model & Parameters & 2p & 3p & 4p & 5p & 2p $\mathcal{L}_S$ \\
\midrule
CNN & 70.6M & 0.0184 / 0.0156 & -- & -- & -- & 0.0117 \\
NMD, main model & 14.5M & 0.0281 / 0.0221 & 0.0405 / 0.0258 & 0.0433 / 0.0250 & 0.0445 / 0.0245 & 0.0180 \\
NMD, improved & 15.0M & 0.0112 / 0.0072 & 0.0321 / 0.0157 & 0.0354 / 0.0154 & 0.0391 / 0.0166 & 0.0071 \\
\bottomrule
\end{tabular}
\caption{CNN baseline \citep{karahan2024deepmultiport} and NMD on the test sets of Table~\ref{tab:error-dist}. Entries are mean / median $S$-parameter MAE. The last column is the training loss $\mathcal{L}_S$ (Appendix~\ref{app:gauge}) on the 2-port set, which averages the absolute errors of the real and imaginary parts instead of the complex modulus and is therefore smaller. The 3-, 4-, and 5-port results of NMD are zero-shot.}
\label{tab:cnn}
\end{table}

\section{RLC network benchmark}
\label{app:rlc}

\paragraph{Circuits.} Each circuit has $M = 8$ nodes. Every node is connected to ground by a capacitor and a resistor, and 10 inductors connect node pairs. Seven of them form a random spanning tree that connects all nodes, and the other 3 connect further random node pairs. Component values are sampled log-uniformly with $C \in [0.5, 2.0]$, $L \in [0.5, 2.0]$, and $R \in [5, 50]$ in normalized units. The first $P = 6$ nodes are port candidates. Responses are evaluated at 36 frequencies uniformly spaced over the normalized angular-frequency band $[0.5, 2.0]$. The exact poles are the eigenvalues of the circuit's state matrix.

\paragraph{Data.} The training set has 48{,}000 circuits, each observed through 3 of its 15 possible 2-port pairs (144{,}000 2-port observations), with validation and test sets of 2{,}000 circuits each. The 2-port test set contains three pairs per test circuit (6{,}000 responses), and each test circuit additionally has one 3-, 4-, and 5-port configuration.

\paragraph{Model.} The circuit is tokenized as a graph with 8 node tokens carrying $(\log C_i, \log R_i)$ and $\binom{8}{2} = 28$ edge tokens carrying the presence of an inductor and its inverse inductance $1/L$, with node embeddings shared between a node token and the edge tokens incident to it so that self-attention acts as message passing. Five transformer layers ($D = 256$) produce the tokens consumed by PoleNet (3 blocks, $K = 24$) and AmpNet (4 blocks), whose port query is a learned embedding of the port node. The loss and weights are those of the RF model ($\lambda_Z = 0.1$, $\lambda_g = 0.5$). Training uses AdamW (learning rate $5 \times 10^{-4}$, weight decay $10^{-3}$, batch size 256, dropout 0.05) for 1{,}000 epochs, with a node-relabeling augmentation that permutes port nodes among themselves and non-port nodes among themselves, which leaves the response invariant. Results are from a single training seed.

\begin{table}[!htbp]
\centering
\small
\begin{tabular}{lccc}
\toprule
Evaluation & $n$ & MAE mean & MAE median \\
\midrule
2p & 6{,}000 & 0.0178 & 0.0164 \\
3p & 2{,}000 & 0.0157 & 0.0144 \\
4p & 2{,}000 & 0.0141 & 0.0131 \\
5p & 2{,}000 & 0.0129 & 0.0119 \\
\bottomrule
\end{tabular}
\caption{$S$-parameter MAE on RLC networks of a model trained only on 2-port responses, evaluated on 2-port test responses and zero-shot on 3-, 4-, and 5-port responses of held-out circuits.}
\label{tab:rlc}
\end{table}

\begin{figure}[!htbp]
  \centering
  \includegraphics[width=\textwidth]{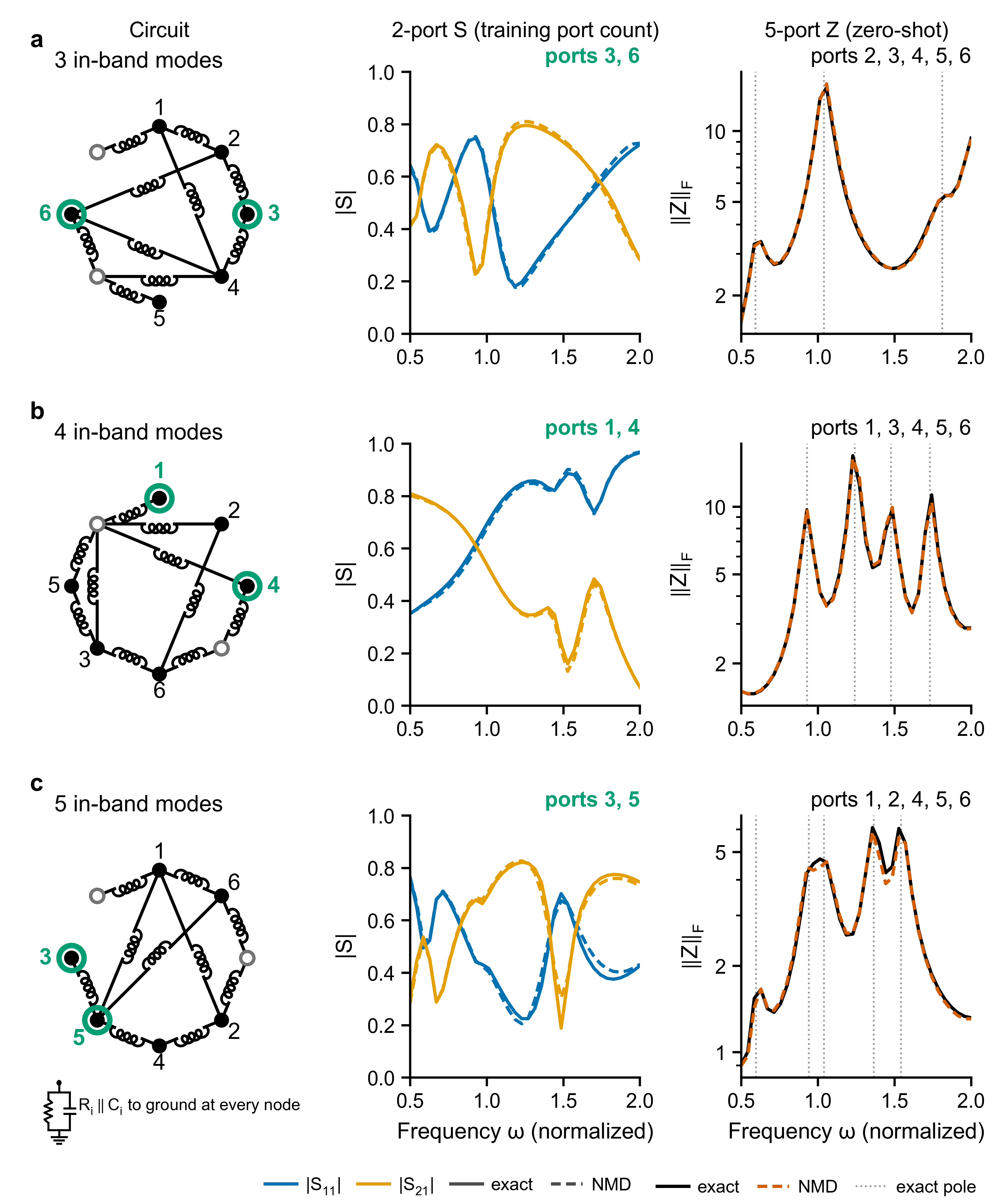}
  \caption{\textbf{RLC test circuits.} (a--c) Circuits with 3, 4, and 5 in-band modes. The left column shows the circuit topology, with port-candidate nodes (filled, numbered), internal nodes (open), an inductor on every edge, and a shunt $R_i \parallel C_i$ to ground at every node (key at bottom left). The ports of the 2-port response in the middle column are circled. The middle column shows $|S_{11}|$ and $|S_{21}|$ of that 2-port response, exact (solid) and NMD (dashed). The right column shows $\|Z(\omega)\|_F$ of the circuit's 5-port test configuration, predicted zero-shot by the 2-port-trained model, with the exact in-band pole frequencies dotted.}
  \label{fig:rlc-circuits}
\end{figure}

\begin{figure}[!htbp]
  \centering
  \includegraphics[width=0.85\textwidth]{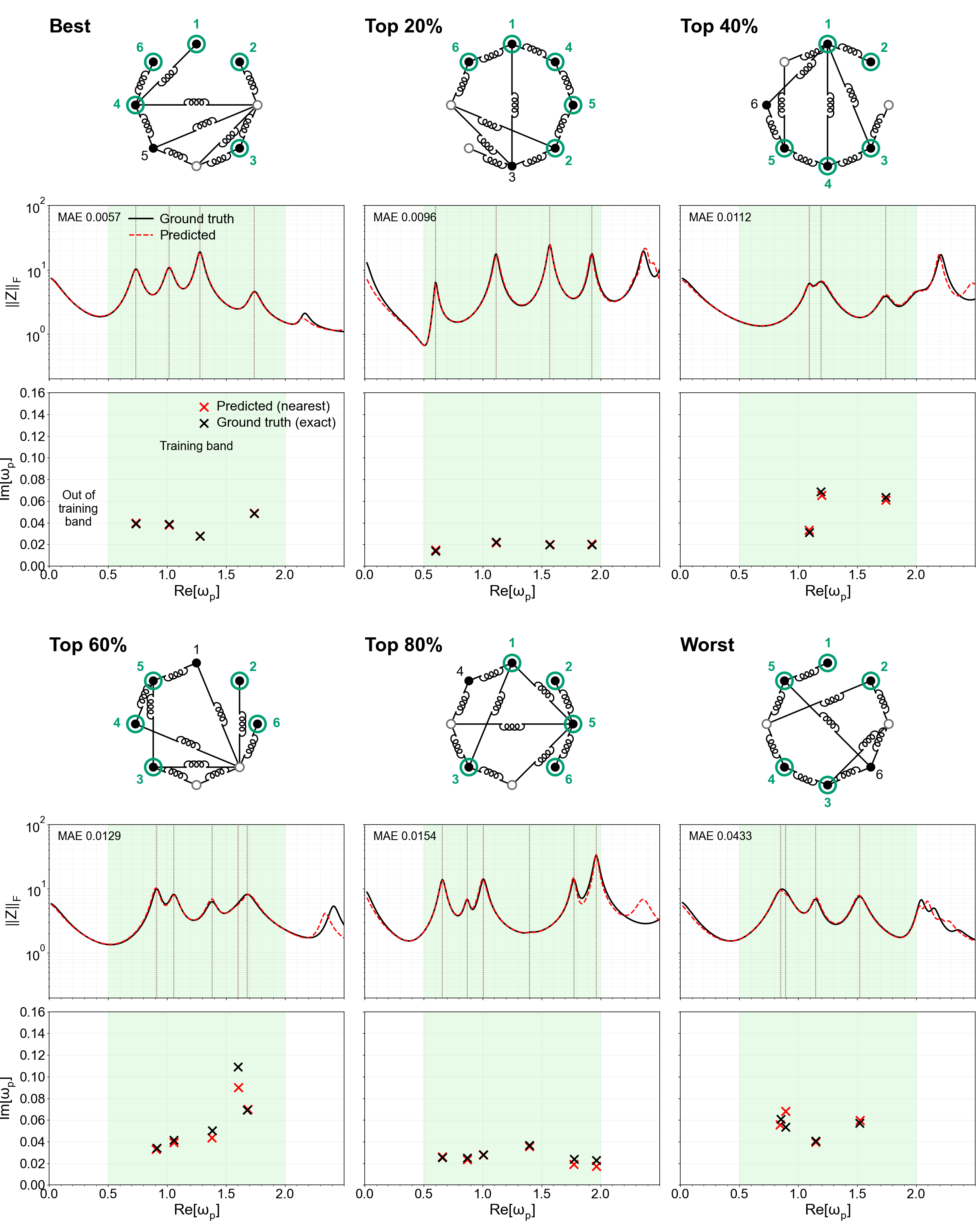}
  \caption{\textbf{Network poles against exact poles}, in the format of Figure~\ref{fig:aaa}. Six test circuits chosen across the range of the 5-port $S$-parameter MAE, from the best to the worst of the 2{,}000 test circuits. ``Top 20\%'' marks the circuit whose error is lower than that of 80\% of the test circuits, and similarly for the other labels. The top row of each block shows the circuit, drawn as in Figure~\ref{fig:rlc-circuits}, with the five ports of its 5-port test configuration circled. The middle row shows $\|Z(\omega)\|_F$ of that 5-port configuration, ground truth (solid) and zero-shot prediction (dashed), evaluated on a dense sweep that extends beyond the training band (shaded), and dotted lines mark the exact and matched pole frequencies. The bottom row shows the poles in the complex frequency plane. Black crosses are the exact in-band poles $\mathrm{eig}(A)$, and red crosses the network pole nearest to each of them among the $K = 24$ predicted poles. The red crosses are selected using the exact poles, so they show that a network pole lies at each physical pole, the criterion behind the recovery rates in Section~\ref{sec:rlc}, not which network poles the model ranks as dominant.}
  \label{fig:rlc-poles}
\end{figure}

\paragraph{Examples.} Figure~\ref{fig:rlc-circuits} shows three test circuits with the 2-port responses the model is trained on and the zero-shot 5-port response, and Figure~\ref{fig:rlc-poles} compares the network poles with the exact poles for six test circuits chosen across the range of the 5-port error, from the best to the worst circuit. In every case the 2-port pair and the 5-port configuration are those of the test sets in Table~\ref{tab:rlc}.

\section{Port-pair holdout and training-set size}
\label{app:pairs}

Every port position appears in training. This is necessary, because a port that is never observed carries no information about how the modes couple to it. Individual port pairs, however, need not appear. To test whether the model memorizes pairs, we removed from training every sample that uses any of 54 selected top--bottom pairs ($12.5\%$ of the training set), together with their mirror images under the flip augmentation, and retrained from scratch (Model C). A control was trained on a random subset of the same size with all pairs present (Model B). The experiment was repeated with 82 held-out pairs ($20.5\%$ of the training set, Models C2 and B2).

Table~\ref{tab:pairholdout} shows that the holdout models predict the held-out pairs as accurately as the seen pairs and match their size-matched controls. The same holds for held-out pairs inside the zero-shot $N$-port matrices, with held-out versus seen MAE of $0.0440$ vs.\ $0.0438$ for Model C and $0.0445$ vs.\ $0.0453$ for Model C2. Consistently, in the full training set the number of samples per top--bottom pair varies by a factor of 168, yet the most frequently seen quarter of the pairs is not more accurate than the least frequently seen quarter (2-port MAE $0.0294$ vs.\ $0.0274$).

\paragraph{Training-set size.} The random-subset controls also show how accuracy depends on the amount of training data. Removing $12.5\%$ and $20.5\%$ of the training data at random raises the 2-port MAE on seen pairs from $0.0293$ to $0.0304$ and $0.0328$ (Models B and B2), a gradual change with no abrupt loss of accuracy.

\begin{table}[!htbp]
\centering
\small
\begin{tabular}{llcc}
\toprule
Model & Training data & 2p seen pairs & 2p held-out pairs \\
\midrule
A & full & 0.0293 & 0.0295 \\
B & random $12.5\%$ removed & 0.0304 & 0.0313 \\
C & 54 pairs held out ($12.5\%$) & 0.0302 & 0.0302 \\
\midrule
B2 & random $20.5\%$ removed & 0.0328 & 0.0323 \\
C2 & 82 pairs held out ($20.5\%$) & 0.0324 & 0.0329 \\
\bottomrule
\end{tabular}
\caption{Port-pair holdout. 2-port test MAE on samples whose port pair is seen or held out in Models C and C2. Models A, B, and B2 saw all pairs, and for them the columns only split the test samples in the same way.}
\label{tab:pairholdout}
\end{table}

\section{Ports on all four edges}
\label{app:lr}

In the main dataset the ports lie on the top and bottom edges only. This is a property of the dataset, not of the architecture, since the left and right edge pixels are formal ports in every EMX simulation and are shorted only in post-processing. We extended the dataset so that 2-port pairs appear on all four edges and retrained the model, still supervised only by 2-port responses. This model also uses an improved training configuration, so its numbers are not directly comparable to Table~\ref{tab:error-dist}. Table~\ref{tab:lr} evaluates zero-shot assembly on configurations that mix ports from different edges, each compared against a separate EMX simulation with port metal present at every queried location. The median MAE stays between $0.010$ and $0.013$ from 3 to 9 ports.

\begin{table}[!htbp]
\centering
\small
\begin{tabular}{lccc}
\toprule
Ports (zero-shot, mixed edges) & $n$ & MAE median & MAE p90 \\
\midrule
3p & 236 & 0.0116 & 0.029 \\
4p to 7p & 416 & 0.0099 to 0.0118 & 0.025 to 0.035 \\
8p to 9p & 112 & 0.0130 to 0.0131 & 0.030 to 0.042 \\
\bottomrule
\end{tabular}
\caption{Zero-shot port-count extension with ports on all four edges.}
\label{tab:lr}
\end{table}

\section{Seed dependency}
\label{app:seeds}

We retrained the main configuration from scratch with 10 different random seeds. The best validation $\mathcal{L}_S$ has mean $0.01831$ and standard deviation $0.00054$ (coefficient of variation $2.9\%$, range $0.01766$--$0.01961$), with no failed or divergent runs. The reported model ($0.01804$) lies within this distribution, $0.5$ standard deviations below the mean.

\section{Broader Impacts}
\label{app:broaderimpacts}
This work contributes a surrogate modeling framework for multi-port linear time-invariant systems, demonstrated here in radio-frequency hardware design, with potential applications in photonic and superconducting quantum hardware design. Positive impacts include faster device design iteration and reduced computational cost of full-wave electromagnetic simulation. We do not foresee specific negative societal impacts beyond those generic to any hardware design tool. The framework operates on physical system observables and does not process human data.


\end{document}